\documentclass[twocolumn]{aastex63}

\usepackage{textcomp, gensymb, amsmath}

\shorttitle{The Upper Edge of the Neptune Desert Is Stable Against Photoevaporation}
\shortauthors{Vissapragada et al.}

\begin{document}

\title{The Upper Edge of the Neptune Desert Is Stable Against Photoevaporation}

\correspondingauthor{Shreyas~Vissapragada}
\email{svissapr@caltech.edu}

\author[0000-0003-2527-1475]{Shreyas~Vissapragada}
\affiliation{Division of Geological and Planetary Sciences, California Institute of Technology, 1200 East California Blvd, Pasadena, CA 91125, USA}

\author[0000-0002-5375-4725]{Heather~A.~Knutson}
\affil{Division of Geological and Planetary Sciences, California Institute of Technology, 1200 East California Blvd, Pasadena, CA 91125, USA}

\author[0000-0002-0371-1647]{Michael~Greklek-McKeon}
\affil{Division of Geological and Planetary Sciences, California Institute of Technology, 1200 East California Blvd, Pasadena, CA 91125, USA}

\author[0000-0002-9584-6476]{Antonija~Oklop{\v{c}}i{\'c}}
\affil{Anton Pannekoek Institute of Astronomy, University of Amsterdam, Science Park 904, 1098 XH Amsterdam, Netherlands}

\author[0000-0002-8958-0683]{Fei~Dai}
\affil{Division of Geological and Planetary Sciences, California Institute of Technology, 1200 East California Blvd, Pasadena, CA 91125, USA}

\author[0000-0002-2248-3838]{Leonardo~A.~dos~Santos}
\affil{Space Telescope Science Institute, 3700 San Martin Drive, Baltimore, MD 21218, USA}

\author[0000-0001-5213-6207]{Nemanja~Jovanovic}
\affil{Department of Astronomy, California Institute of Technology, 1200 East California Blvd, Pasadena, CA 91125, USA}

\author[0000-0002-8895-4735]{Dimitri~Mawet}
\affil{Department of Astronomy, California Institute of Technology, 1200 East California Blvd, Pasadena, CA 91125, USA}
\affil{Jet Propulsion Laboratory, California Institute of Technology, 4800 Oak Grove Dr, Pasadena, CA 91109, USA}

\author[0000-0001-6205-9233]{Maxwell~A.~Millar-Blanchaer}
\affil{Department of Physics, University of California, Santa Barbara, CA 93106, USA}

\author[0000-0003-0062-1168]{Kimberly~Paragas}
\affil{Division of Geological and Planetary Sciences, California Institute of Technology, 1200 East California Blvd, Pasadena, CA 91125, USA}

\author[0000-0002-5547-3775]{Jessica~J.~Spake}
\affil{Division of Geological and Planetary Sciences, California Institute of Technology, 1200 East California Blvd, Pasadena, CA 91125, USA}

\author[0000-0002-1481-4676]{Samaporn~Tinyanont}
\affiliation{Department of Astronomy and Astrophysics, University of California, Santa Cruz, CA 95064, USA}

\author[0000-0002-1871-6264]{Gautam~Vasisht}
\affil{Jet Propulsion Laboratory, California Institute of Technology, 4800 Oak Grove Dr, Pasadena, CA 91109, USA}

\begin{abstract}
Transit surveys indicate that there is a deficit of Neptune-sized planets on close-in orbits. If this ``Neptune desert'' is entirely cleared out by atmospheric mass loss, then planets at its upper edge should only be marginally stable against photoevaporation, exhibiting strong outflow signatures in tracers like the metastable helium triplet. We test this hypothesis by carrying out a 12-night photometric survey of the metastable helium feature with Palomar/WIRC, targeting seven gas-giant planets orbiting K-type host stars. Eight nights of data are analyzed here for the first time along with reanalyses of four previously-published datasets. We strongly detect helium absorption signals for WASP-69b, HAT-P-18b, and HAT-P-26b; tentatively detect signals for WASP-52b and NGTS-5b; and do not detect signals for WASP-177b and WASP-80b. We interpret these measured excess absorption signals using grids of Parker wind models to derive mass-loss rates, which are in good agreement with predictions from the hydrodynamical outflow code ATES for all planets except WASP-52b and WASP-80b, where our data suggest that the outflows are much smaller than predicted. Excluding these two planets, the outflows for the rest of the sample are consistent with a mean energy-limited outflow efficiency of $\varepsilon = 0.41^{+0.16}_{-0.13}$. Even when we make the relatively conservative assumption that gas-giant planets experience energy-limited outflows at this efficiency for their entire lives, photoevaporation would still be too inefficient to carve the upper boundary of the Neptune desert. We conclude that this feature of the exoplanet population is a pristine tracer of giant planet formation and migration mechanisms.
\end{abstract}

\section{Introduction} \label{sec:intro}
\begin{figure}[ht]
\centering
\includegraphics[width=0.5\textwidth]{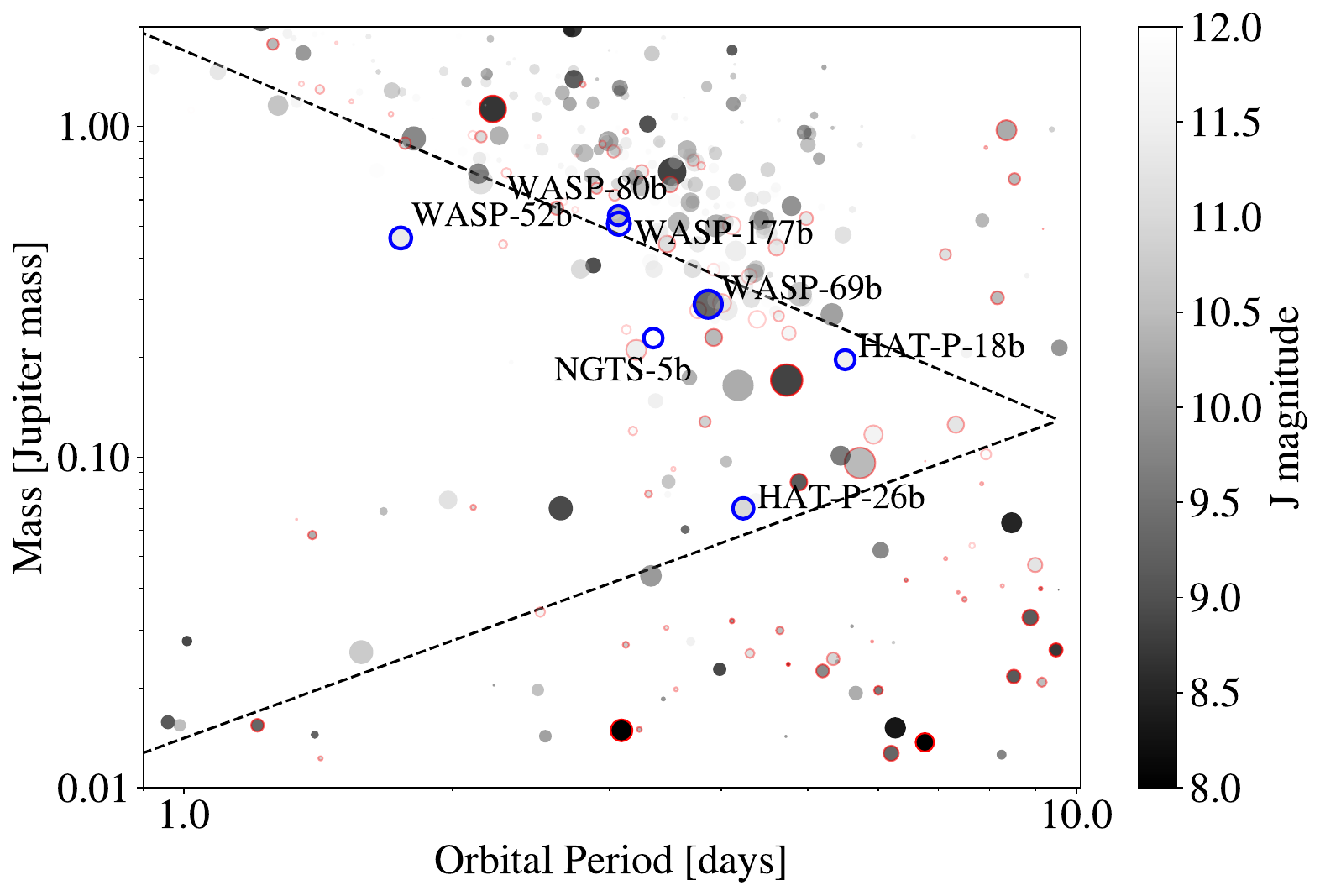}
\caption{Transiting exoplanets with fractional mass uncertainties of less than 30\%, with the dashed lines indicating the Neptune desert boundaries from \citet{Mazeh16}. Planets orbiting stars with $4000$~K$<T_\mathrm{eff}<5400$~K are outlined in red, and the seven targets constituting our sample are outlined in blue. The point color denotes the host star's J magnitude, and the point size is proportional to logarithm of the predicted SNR in the WIRC helium bandpass from Section~\ref{sec:sample}.}
\label{fig:desert}
\end{figure}

The transit and radial velocity techniques have revealed masses and radii for thousands of planets spanning a wide range of orbital periods. These surveys indicate that relatively few sub-Jovian planets reside on close-in orbits ($P \lesssim 5$ days). This ``Neptune desert'' is a robust feature of the planetary census, which is relatively complete in this regime \citep{Szabo11, Beauge13, Lundkvist16}. \citet{Mazeh16} defined boundaries for the Neptune desert by searching for curves that maximize the contrast between regions of the mass-period plane with and without planets. We plot their definition for the Neptune desert in Figure~\ref{fig:desert} along with the planetary population observed to date.

Given its close-in location, it is natural to wonder whether the desert might have been created by atmospheric mass loss \citep[e.g.][]{Youdin11, Beauge13, Kurokawa14, Ionov18, Owen18}. Planets with short orbital periods are subjected to intense amounts of high-energy radiation that can drive atmospheric outflows \citep{Owen19}. The resulting atmospheric mass-loss rates can be large, especially for planets orbiting young stars, which emit proportionally more of their energy in high-energy photons. If close-in gas-giant planets formed \textit{in situ} \citep[e.g.][]{Batygin16} or arrived at their present-day locations relatively early ($\sim$10~Myr) via disk migration \citep[e.g.][]{Ida08}, they might have experienced a period of strong photoevaporation \citep[e.g.][]{MurrayClay09}. Hot Neptunes typically have lower gravitational potentials than their Jovian counterparts, making them more susceptible to photoevaporation and/or Roche lobe overflow \citep{Kurokawa14, Valsecchi15, Koskinen22}. This means that Jupiters could survive at orbital separations where most Neptunes would be destroyed. 

If this population of close-in giant planets instead arrived at their present locations relatively late ($\sim$Gyr) via dynamical interactions with other bodies in their systems \citep{Rasio96}, we would expect photoevaporation to play a relatively minor role in their long-term evolution. In this scenario, planets undergoing high-eccentricity migration circularized onto orbits with final semimajor axes of $a \approx 2r_\mathrm{peri}$, where $r_\mathrm{peri}$ is the pericenter distance when the eccentricity $e\sim1$. Neptune-mass planets with small present-day semimajor axes would have had pericenter passages inside the stellar tidal disruption radius, whereas Jupiter-mass planets would survive at comparable separations \citep{Ford06, Guillochon11, Matsakos16, Owen18}.

Our ability to explain the Neptune desert is thus intimately linked to our understanding of the origins of close-in giant planets \citep{Dawson18}. Unfortunately, previous studies have found it difficult to differentiate between atmospheric escape and high-eccentricity migration as explanations for this feature. Depending on their assumptions about the energy-limited mass-loss efficiency, stellar X-ray and ultraviolet (XUV) flux, initial star-planet separation, and pressure at the XUV photobase, some published studies suggest that sub-Jovian planets near the upper edge of the desert are stable against evaporation, while other suggest that they experience runaway envelope escape \citep{Kurokawa14, Ionov18, Owen18, Rao21}. High-eccentricity migration models also have their own uncertainties, most notably the assumed tidal quality factor, the initial star-planet separation, and the eccentricity excitation mechanism \citep{Matsakos16, Owen18}. Previous observational studies have used the orbital properties and multiplicity of hot Jupiter systems to constrain the fraction of hot Jupiters that might have formed via high-eccentricity migration \citep[e.g.][]{Winn10, Knutson14, Dawson15, Ngo16, Rice22}. Although it appears unlikely that all hot Jupiters formed this way, such planets almost certainly comprise a subset of the hot Jupiter population \citep[see the review by][]{Dawson18}.

Observations of atmospheric escape can provide a complementary basis to differentiate between these scenarios for the origin of the Neptune desert. If atmospheric escape is indeed the mechanism that clears the Neptune desert, planets at its edge would be marginally stable against photoevaporation, potentially exhibiting large outflows today. Recent studies have demonstrated that the metastable helium triplet can be used to detect and characterize planetary outflows for close-in transiting planets \citep[e.g.][]{Spake18, Oklopcic18}. Unlike the Lyman-$\alpha$ transit depth -- which is controlled not by the planetary mass-loss rate, but by the distance neutral hydrogen atoms can travel before they are photoionized \citep{Owen21} -- the metastable helium absorption amplitude probes low-velocity thermospheric gas near the wind-launching radius. This means that it can be readily modeled using a one-dimensional isothermal Parker wind to infer mass-loss rates as low as $10^{9}-10^{10}$~g~s$^{-1}$ \citep[e.g.][]{Mansfield18, Oklopcic18, Lampon20, Vissapragada20b}. For comparison, the required average mass-loss rate to clear a hypothetical $M_\mathrm{J}$ planet from the desert in a few hundred Myr is of order $10^{14}$~g~s$^{-1}$. The stellar XUV luminosity (on which the planetary mass-loss rate is linearly dependent in the energy limit) decreases by about 2 orders of magnitude over the stellar lifetime \citep[e.g.][]{Johnstone21}. Therefore, if planets at the upper edge of the desert are only marginally stable against photoevaporation, they would be expected to exhibit relatively strong outflow signatures even given the reduced XUV fluxes of their middle-aged host stars. Although uncertainties in the assumed incident XUV spectrum, outflow temperature, and outflow composition limit our ability to predict the metastable helium population in the Parker wind model, we can still constrain present-day mass-loss rates well enough for planetary evolution studies. This is especially true when we incorporate additional information like line shapes and energetics \citep{dosSantos22, Vissapragada22}. 

We recently installed a narrowband filter centered on the metastable helium line \citep{Vissapragada20b} in the Wide-field InfraRed Camera \citep[WIRC;][]{Wilson03} at the Hale 200-inch Telescope at Palomar Observatory. We commissioned this new observing mode by confirming the outflow of WASP-69b, finding an absorption signal commensurate with spectroscopically resolved observations by \citet{Nortmann18}. As part of this same study we also observed a transit of WASP-52b but did not detect a strong signal. We have since used this same observing mode to detect an atmospheric outflow from HAT-P-18b \citep{Paragas21} and found a tentative evidence for an outflow from the young planet V1298 Tau d \citep{Vissapragada21}. 

From 2019-2021, we used our narrowband helium photometer to survey the metastable helium line in a seven-planet sample near the upper edge of the Neptune desert. There are currently ten published helium detections in the literature: WASP-107b \citep{Spake18, Allart19, Kirk20, Spake21}, HAT-P-11b \citep{Allart18, Mansfield18}, WASP-69b \citep{Nortmann18, Vissapragada20b}, HD 209458b \citep{AlonsoFloriano19}, HD 189733b \citep{Salz18, Guilluy20}, GJ 3470b \citep{Palle20, Ninan20}, HAT-P-18b \citep{Paragas21}, HAT-P-32b \citep{Czesla22}, TOI-560.01 \citep{Zhang22}, and GJ 1214b \citep{OrellMiquel22}. These observations, along with many non-detections and tentative detections in the literature, span wide ranges in planetary mass and radius, stellar spectral type and activity level, system age, observing methodology, and data reduction technique. This heterogeneity can obfuscate underlying trends in the data, because the measured helium absorption signal depends on a variety of factors that all vary across the sample. We therefore focus on a uniformly selected sample of helium observations obtained with Palomar to search for trends in measured mass-loss rates for planets at the upper edge of the Neptune desert.

In this work we present the results of our survey, including a re-analysis of the published transit observations from \cite{Vissapragada20b} and \cite{Paragas21}. We also present new observations of eight additional transits for five planets: WASP-52b, WASP-80b, WASP-177b, HAT-P-26b, and NGTS-5b. In Section~\ref{sec:obs}, we describe our sample selection methodology as well as our observation and data reduction procedures. In Section~\ref{sec:lightcurves}, we present helium-band light curves for each of our seven targets. In Section~\ref{sec:massloss}, we model each of our light curves using a one-dimensional isothermal Parker wind model to constrain the planetary mass-loss rates. We discuss the implications of our results for the population-level picture of atmospheric escape in Section~\ref{sec:disc}, and summarize our conclusions in Section~\ref{sec:conc}. 

\section{Observations} \label{sec:obs}
\subsection{Sample Selection} \label{sec:sample}

\begin{deluxetable*}{cccccccccccc}[ht]
\tabletypesize{\scriptsize}
\tablecaption{Stellar and planetary parameters for sample selected in this work \label{table:sample}.}
\tablehead{\colhead{Star}  & \colhead{{\it J} mag}  & \colhead{$T_\star$ (K)} & \colhead{$R_\star$ ($R_\Sun$)} & \colhead{$M_\star$ ($M_\Sun$)} & \colhead{$\log(R_\mathrm{HK}')$} & \colhead{$M_\mathrm{p}$ ($M_\mathrm{J}$)} & \colhead{$R_\mathrm{p}$ ($R_\mathrm{J}$)} & \colhead{$\rho_\mathrm{p}$ (g~cm$^{-3}$)} & \colhead{$P$ (d)} & \colhead{$a$ (au)} & \colhead{Reference}}
\startdata
WASP-69 & 8.0 & 4715$^{+50}_{-50}$ & 0.813$^{+0.028}_{-0.028}$ & 0.826$^{+0.029}_{-0.029}$ & -4.54 & 0.260$^{+0.017}_{-0.017}$ & 1.057$^{+0.047}_{-0.047}$ & 0.291$^{+0.041}_{-0.041}$ &3.87 & 0.045 & A14 \\
WASP-52 & 10.6 & 5000$^{+100}_{-100}$ & 0.79$^{+0.02}_{-0.02}$ & 0.87$^{+0.03}_{-0.03}$ & -4.4 & 0.459$^{+0.022}_{-0.021}$ & 1.27$^{+0.03}_{-0.03}$ & 0.29$^{+0.03}_{-0.03}$ & 1.75 & 0.027 & H13 \\
HAT-P-18 & 10.8 & 4803$^{+80}_{-80}$ & 0.749$^{+0.037}_{-0.037}$ & 0.770$^{+0.031}_{-0.031}$ & -4.73 & 0.197$^{+0.013}_{-0.013}$ & 0.995$^{+0.052}_{-0.052}$ & 0.25$^{+0.04}_{-0.04}$ & 5.51 & 0.056 & H11a, V22 \\
WASP-80 & 9.2 & 4143$^{+92}_{-94}$ & 0.586$^{+0.017}_{-0.018}$ & 0.577$^{+0.051}_{-0.054}$ & -4.04 & 0.538$^{+0.035}_{-0.036}$ & 0.999$^{+0.030}_{-0.031}$ & 0.717$^{+0.039}_{-0.032}$ & 3.07 & 0.034 & T15, F21 \\
WASP-177 & 10.7 & 5017$^{+70}_{-70}$ & 0.885$^{+0.046}_{-0.046}$ & 0.876$^{+0.038}_{-0.038}$ & -- & 0.508$^{+0.038}_{-0.038}$ & 1.58$^{+0.66}_{-0.36}$ & 0.172$^{+0.203}_{-0.110}$ & 3.07 & 0.040 & T19 \\
HAT-P-26 & 10.1 & 5079$^{+88}_{-88}$ & 0.788$^{+0.098}_{-0.043}$ & 0.816$^{+0.033}_{-0.033}$ & -4.99 & 0.059$^{+0.007}_{-0.007}$ & 0.565$^{+0.072}_{-0.032}$ & 0.40$^{+0.10}_{-0.10}$ & 4.23 & 0.048 & H11b \\
NGTS-5 & 12.1 & 4987$^{+41}_{-41}$ & 0.739$^{+0.014}_{-0.012}$ & 0.661$^{+0.068}_{-0.061}$ & -4.63 & 0.229$^{+0.037}_{-0.037}$ & 1.136$^{+0.023}_{-0.023}$ & 0.193$^{+0.032}_{-0.034}$ & 3.36 & 0.038 & E19 
\enddata
\tablecomments{{\it J} magnitudes are taken from 2MASS \citep{Cutri03}. Orbital periods and semimajor axes are truncated with errors omitted for clarity. In the references column, A14 is \citet{Anderson14}, H13 is \citet{Hebrard13}, H11a is \citet{Hartman11a}, V22 is \citet{Vissapragada22}, T15 is \citet{Triaud15}, F21 is \citet{Fossati22}, T19 is \citet{Turner19}, H11b is \citet{Hartman11b}, and E19 is \citet{Eigmuller19}. The $\log(R_\mathrm{HK}')$ value for NGTS-5b was calculated using an S-value from P. Eigm\"{u}ller (private communication).}
\end{deluxetable*}

To construct our survey sample, we searched the NASA Exoplanet Archive \citep{Akeson13, ps} for transiting giant ($R > 4 R_\Earth$) planets with measured masses orbiting K-type host stars ($T_\star = 4000-5400$~K). Planets orbiting K-type stars are predicted to have the largest fractional metastable helium populations, and therefore to exhibit the strongest helium absorption signatures during transit \citep{Oklopcic19, Wang21a}. This is because K stars have relatively high EUV fluxes, which populate the metastable state via ground-state ionization and subsequent recombination, and relatively low mid-UV fluxes, which ionize and depopulate the metastable state. A large fraction of the EUV flux also comes from coronal iron lines, and \citet{Poppenhaeger22} highlighted that this consideration also favors K stars for helium studies. The stellar spectral type gives only an approximate expectation for the high-energy spectrum of the star however, and the discovery of a strong helium signature for HAT-P-32b (a gas-giant planet orbiting an F star) is a reminder that other stars with favorable spectral energy distributions (SEDs) can also exhibit strong signals regardless of their spectral type \citep{Czesla22}. Nevertheless, the bulk of planets with detected outflows in the literature orbit K stars.

We used this initial sample to create a rank-ordered list based on predicted signal-to-noise ratio (SNR). We estimated the SNR by first calculating lower atmospheric planetary scale heights using masses, radii, and equilibrium temperatures from the NASA Exoplanet Archive with a mean molecular weight of 2.3 amu. We then assumed that a typical helium absorption extends 85.5 lower atmospheric scale heights at the core of the helium feature, corresponding to the magnitude of the measured excess absorption signal for WASP-69b \citep{Nortmann18}. We selected this planet as a benchmark simply because it was among the first published measurements of a planetary helium line at high resolving power, but a different choice for the signal amplitude would not have affected the relative rankings. Assuming a line shape similar to that observed by \citet{Nortmann18}, we convolved the expected absorption features with our measured filter transmission profile to get a Palomar/WIRC signal estimate. We calculated the expected photometric noise for each observation by scaling on-sky noise statistics from previous observations \citep{Vissapragada20a, Vissapragada20b} to the 2MASS \textit{J} magnitudes for each host star. From the resulting rank-ordered list, we removed two targets with extensive previous observations \citep[WASP-107b and HD 189733b;][]{Salz18, Spake18, Allart19, Guilluy20, Kirk20, Spake21} and one target that lacked nearby comparison stars, which are required for our photometric observation technique, and had an exceptionally long transit duration (KELT-11b). 

We were able to obtain observing time for eight of the highest-priority targets with predicted SNR $> 3$, one of which (HAT-P-12b) we did not ultimately observe due to telescope closures and poor weather conditions. In Table~\ref{table:sample}, we present the stellar and planetary parameters for the remaining seven targets constituting our sample. All of the new targets in our sample except for NGTS-5b (a relatively recent discovery) were independently identified by \citet{Kirk20} as good candidates for metastable helium observations. These seven planets lie near the edge of the Neptune desert, visualized in Figure~\ref{fig:desert}. HAT-P-26b is close to the lower edge of the desert, and the other six planets trace the upper edge.

\subsection{Palomar/WIRC Observations}

\begin{deluxetable*}{cccccccccccc}[t!]
\tabletypesize{\scriptsize}
\tablecaption{Summary of Palomar/WIRC observations analyzed in this work \label{table:log}.}
\tablehead{\colhead{Planet}  & \colhead{Date}  & \colhead{Start Time} & \colhead{End Time} & \colhead{$n_\mathrm{exp}$} & \colhead{$t_\mathrm{exp}$ (s) } & \colhead{Start/Min/End Airmass} & \colhead{$n_\mathrm{comp}$} & \colhead{$n_\mathrm{dither}$} & \colhead{$r_\mathrm{phot}$} (px) & \colhead{$\sigma$ [\%]} & \colhead{$\sigma/\sigma_\mathrm{phot}$}}
\startdata
WASP-69b & 2019 Aug 16 & 04:26:06 & 11:01:00 & 345 & 60 & 1.73/1.28/2.52 & 4 & 4 & 10 & 0.24 & 2.2 \\
WASP-52b & 2019 Sep 17 & 03:16:57 & 11:14:49 & 291 & 90 & 2.04/1.10/1.97 & 4 & 4 & 7 & 0.70 & 1.4 \\
HAT-P-18b & 2020 Jun 05 & 05:03:30 & 10:41:41 & 207 & 90 & 1.24/1.00/1.21 & 4 & 4 & 7 & 1.15 & 1.9 \\
HAT-P-18b & 2020 Jul 08 & 05:06:57 & 11:03:26 & 217 & 90 & 1.01/1.00/2.33 & 6 & 4 & 11 & 0.82 & 1.2 \\
WASP-80b & 2020 Jul 09 & 05:20:35 & 11:28:27 & 311 & 60 & 2.05/1.23/1.61 & 5 & 26 & 8 & 0.56/0.54 & 2.3/2.1 \\
WASP-52b & 2020 Aug 04 & 05:48:11 & 11:03:11 & 192 & 90 & 2.33/1.10/1.13 & 3 & 6 & 11 & 0.87 & 1.4 \\
WASP-177b & 2020 Oct 02 & 03:01:32 & 08:53:57 & 215 & 90 & 1.50/1.22/2.06 & 2 & 4 & 9 & 0.73 & 1.2\\
HAT-P-26b & 2021 Feb 18 & 08:19:21 & 12:48:06 & 165 & 90 & 2.03/1.15/1.17 & 1 & 4 & 12 & 1.80 & 2.8\\
NGTS-5b & 2021 Apr 30 & 05:56:31 & 11:31:18 & 204 & 90 & 1.30/1.13/1.82 & 1 & 7 & 8 & 3.80 & 1.7 \\
HAT-P-26b & 2021 May 01 & 04:34:13 & 11:40:54 & 260 & 90 & 1.53/1.15/2.53 & 2 & 5 & 12 & 0.81 & 1.6 \\
HAT-P-26b & 2021 May 18 & 04:20:36 & 10:29:36 & 225 & 90 & 1.30/1.15/2.44 & 2 & 7 & 9 & 0.61 & 1.4 \\
NGTS-5b & 2021 May 27 & 04:25:11 & 10:20:52 & 217 & 90 & 1.26/1.13/2.27 & 1 & 7 & 8 & 2.5 & 1.5 \\
\enddata
\tablecomments{In the column headers, $n_\mathrm{exp}$ refers to the total number of exposures, $t_\mathrm{exp}$ is the exposure time, $n_\mathrm{comp}$ is the number of comparison stars, $n_\mathrm{dither}$ is the total number of dither frames used in constructing the background, $r_\mathrm{phot}$ is the radius of the circular aperture used in the photometric extraction, $\sigma$ is the rms scatter of the residuals to our final light-curve fit, and $\sigma/\sigma_\mathrm{phot}$ is the ratio of the rms scatter to the photon noise limit. The two $\sigma$ and $\sigma/\sigma_\mathrm{phot}$ values for the WASP-80b observation indicate the noise before and after the instrument was reset. All dates and times are UT.}
\end{deluxetable*}

From 2019--2021, we observed transits for the seven targets in our sample over 12 nights. We summarize these observations in Table~\ref{table:log}. All transits were observed in a custom metastable helium filter centered at 1083.3~nm with a full-width at half-maximum (FWHM) of 0.635~nm, previously introduced in \citet{Vissapragada20b}. Each observation followed a similar sequence to that described in our previous papers \citep{Vissapragada20b, Paragas21, Vissapragada21}. We began each night by observing a helium arc lamp through the helium filter, allowing us to position the target star where the effective transmission function of the filter was centered on the helium feature. The exact positioning within this region was selected to maximize the number of suitable comparison stars; the number of comparison stars for each night is noted in Table~\ref{table:log}. On most nights, after acquiring and positioning the target star we used a custom near-infrared (NIR) beam-shaping diffuser to mold the PSF into a top-hat 3$\arcsec$ in diameter, allowing for precise control of time-correlated systematics \citep{Stefansson17, Vissapragada20a}. The only exception was our observation of HAT-P-18b on UT 2020 Jun 05. Weather conditions were poor that night, and we chose to slightly defocus the telescope to 1$\farcs$2 instead of using the diffuser in order to minimize the sky background flux in our photometric aperture. As described in \cite{Vissapragada20b}, the center wavelength of the filter shifts across the field of view, and the resulting sky background is highly structured with bright rings corresponding to OH emission features. To correct for the structured background, we obtained a number of dithered background frames on each night (noted in Table~\ref{table:log}).
\subsection{Data Reduction} \label{redux}

We dark-corrected and flat-fielded each image, and then corrected detector cosmetics including bad pixels and residual striping \citep[for a more detailed description, see][]{Vissapragada20a}. Because the effective bandpass of the filter changes across the field of view, we also corrected for telluric emission lines and time-varying telluric water absorption. This is necessary because comparison stars occupy different parts of the field and thus experience different amounts of OH emission and time-varying water absorption \citep{Vissapragada20b}. To correct the sky background and telluric emission lines from each science frame, we first constructed a background frame from the aforementioned dither sequence. Then, we sigma-clipped the science frame to remove sources, and finally we median-scaled the dither frame to match the science frame in 10~px radial steps from the ``zero-point'', which is where light encounters the filter at normal incidence \citep{Vissapragada20b}. We demonstrated in \citet{Paragas21} that the scaling factors from this procedure can also be used to decorrelate the time-varying telluric water absorption, which can otherwise contaminate the transit signal. The OH line at 1083.4~nm overlaps with nearby telluric water features at the resolution of our filter, and its flux evolves differently over the night relative to the uncontaminated OH lines. By taking the ratio of scaling factors in the contaminated and uncontaminated OH lines, we can construct a proxy for the time-varying telluric water absorption. This proxy is included as a covariate in our light-curve fits. 

We performed aperture photometry on each source using the \texttt{photutils} package, following the procedure detailed in \citet{Vissapragada20b}. We tested apertures from 3 to 15 pixels (0$\farcs$75--3$\farcs$75) in radius for each transit observation. We allowed the locations of the apertures to shift (separately for each star) from image to image to account for variations in telescope pointing. We found that the pointing variations were smaller than $1\arcsec$ across all datasets except for our observation of WASP-80b on UT 2020 Jul 09, where the instrument crashed during the observation and had to be reset. We included the centroid offsets as covariates in our light-curve fits along with the airmass curves, which we found to be important for a subset of the observations. To select the optimal aperture for use in light-curve modeling, we first normalized the target star's light curve by an average of the comparison star light curves, and subsequently removed 4$\sigma$ outliers using a moving median filter. We then selected the aperture size that minimized the per-point rms scatter in the averaged and moving-median-corrected target star photometry. The optimized aperture sizes are noted in Table~\ref{table:log}.

\subsection{Light-curve Modeling} \label{sec:modeling}
After calculating the optimized photometry for each transit observation along with the corresponding decorrelation vectors (airmass, centroid offset, and absorption proxy), we proceeded to fit the transit light curves using \texttt{exoplanet} \citep{ForemanMackey21a, ForemanMackey21b}. Our procedure is similar to that described in \citet{Paragas21} and \citet{Vissapragada21}. Briefly, each target light curve is modeled with a limb-darkened \texttt{starry} transit light curve \citep{Luger19}, which is multiplied by a systematics model. The transit light curve is parameterized by the radius ratio $R_\mathrm{p}/R_\star$, the orbital period $P$, the epoch $T_0$, the scaled semi-major axis $a/R_\star$, the impact parameter $b$, and the quadratic limb darkening coefficients ($u_1$, $u_2$). 

We experiment with leaving the limb darkening coefficients free 
\citep[sampling them using the approach from][]{Kipping13} and fixing them to values computed from \texttt{ldtk} \citep{Husser13, Parviainen15}. Both approaches have advantages and drawbacks. Using the computed values can be advantageous for faint stars, where the precision on the transit shape is too low for our data to constrain the limb darkening themselves. On the other hand, the \texttt{PHOENIX} models \citep{Husser13} underlying \texttt{ldtk} may not accurately describe the stellar brightness profile in the helium line because the line is formed in the chromosphere. For some stars in our sample, we observe mismatches between the \texttt{ldtk} calculations and the retrieved limb darkening parameters; this could either be due to strong stellar limb darkening or an optically-thin component of the outflow altering the transit shape and biasing the inference for the observed limb darkening parameters \citep[e.g.][]{MacLeod22, Wang21a, Wang21b}. Therefore, we repeat the fits for each planet leaving the limb darkening coefficients free and holding them fixed to the calculated values, and adopt the fixed solution when it agrees with the free retrieval or when the free retrieval is otherwise uninformative on the limb darkening coefficients. We show how the posteriors on the limb darkening coefficients compare to the calculated values in Appendix~\ref{ldcs}.

We fix the eccentricities to zero in our fits; none of the planets in our sample are known to have eccentric orbits. Although we note that there is tentative evidence (1-2$\sigma$) that HAT-P-18b and HAT-P-26b may have small but non-zero orbital eccentricities \citep{Hartman11a, Hartman11b, Wallack19}, the eccentricities are too small to have an impact on the light-curve modeling. We modeled the systematics as a linear combination of comparison star light curves and decorrelation vectors, with the weights left as free parameters in each fit. We also included the mean-subtracted BJD times as a decorrelation vector, which functions as a linear baseline. Each weight was assigned a uniform prior of $\mathcal{U}(-2, 2)$. We list the rest of the priors for the transit light-curve modeling in Table~\ref{table:priors}.

\begin{deluxetable*}{cccccccc}[ht]
\tabletypesize{\scriptsize}
\tablecaption{Priors for the transit light-curve fits. \label{table:priors}}
\tablehead{\colhead{Planet}  & \colhead{$P$ (days)}  & \colhead{$T_0$ (BJD - 2450000)} & \colhead{$b$} & \colhead{$a/R_\star$} & \colhead{WIRC $R_\mathrm{p}/R_\star$} & \colhead{References}}
\startdata
WASP-69b & $\mathcal{N}(3.8681390, 0.0000006)$ & $\mathcal{N}(7176.17789, 0.00017)$ & $\mathcal{N}(0.686, 0.023)$ & $\mathcal{N}(12.00, 0.46)$ & $\mathcal{U}(0., 0.25)$ & A14, K22 \\
WASP-52b & $\mathcal{N}(1.74978119, 0.00000010)$ & $\mathcal{N}(6770.05972, 0.00004)$ & $\mathcal{N}(0.60, 0.02)$ & $\mathcal{N}(7.38, 0.11)$ &  $\mathcal{U}(0., 0.25)$ & H13, K22 \\
HAT-P-18b & $\mathcal{N}(5.5080300, 0.0000008)$ & $\mathcal{N}(7276.25646, 0.00010)$ & $\mathcal{N}(0.352, 0.057)$ & $\mathcal{N}(16.39, 0.24)$ &  $\mathcal{U}(0., 0.25)$ & K17, K22\\
WASP-80b & $\mathcal{N}(3.06785271, 0.00000019)$ & $\mathcal{N}(6671.49615, 0.00004)$ & $\mathcal{N}(0.215, 0.022)$ & $\mathcal{N}(12.63, 0.13)$ &  $\mathcal{U}(0., 0.25)$ & T15, K22 \\
WASP-177b & $\mathcal{N}(3.071722, 0.000001)$ & $\mathcal{N}(7994.37140, 0.00028)$ & $ \mathcal{N}(0.980, 0.092)$ & $\mathcal{N}(9.610, 0.530)$ &  $\mathcal{U}(0., 0.4)$ & T19 \\
HAT-P-26b & $\mathcal{N}(4.2345002, 0.0000007)$ & $\mathcal{N}(6892.59046, 0.00010)$ & $\mathcal{N}(0.303, 0.122)$ & $\mathcal{N}(13.06, 0.83)$ &  $\mathcal{U}(0., 0.25)$ & H11, K22 \\
NGTS-5b & $\mathcal{N}(3.3569866, 0.0000026)$ & $\mathcal{N}(7740.35262, 0.00026)$ & $\mathcal{N}(0.653, 0.048)$ & $\mathcal{N}(11.111, 0.32)$ &  $\mathcal{U}(0., 0.4)$ & E19 \\
\enddata
\tablecomments{$\mathcal{U}(a,b)$ indicates a uniform prior between $a$ and $b$ and $\mathcal{N}(a,b)$ indicates a normal prior with mean $a$ and standard deviation $b$. In the References column, A14 is \citet{Anderson14}, K22 is \citet{Kokori22}, H13 is \citet{Hebrard13}, K17 is \citet{Kirk17}, T15 is \citet{Triaud15}, T19 is \citet{Turner19}, H11 is \citet{Hartman11b}, and E19 is \citet{Eigmuller19}.}
\end{deluxetable*}

After defining the model, we sampled the posterior distributions for each model parameter using the No U-Turn Sampler \citep[NUTS][]{Hoffman11} implemented in \texttt{pymc3} \citep{Salvatier16}. In each light-curve fit, we ran four chains for 1,500 tuning steps each before taking 2,500 draws. This resulted in a total of 10,000 draws, which we used to derive the final results. When performing joint fits across multiple nights of WIRC data, we doubled the number of tuning steps and draws. We verified the convergence of each fit by visually inspecting the trace plots to ensure that the chains were well-mixed, and also by checking that the Gelman-Rubin statistic \citep{Gelman92} was $\hat{R}\ll1.01$ for every sampled parameter. 

In addition to the comparison star photometry, each night of data has three additional covariates as described in Section~\ref{redux}: the airmass curve, the water absorption proxy, and the centroid offsets. We determined which of these covariates to include in the final model by repeating the fit with every combination of these three parameters and calculated the Bayesian Information Criterion \citep[BIC;][]{Schwarz78} for each fit:
\begin{equation}
    \mathrm{BIC} = k\ln{n} - 2\ln{\hat{\mathcal{L}}},
\end{equation}
where $k$ is the number of fit parameters, $n$ is the number of data points, and $\mathcal{\hat{L}}$ is the maximum likelihood estimate, or MLE. We report the differences in BIC ($\Delta$BIC) between each fit and a fit with no additional covariates in Appendix~\ref{modelselection}, and select the model with the minimal BIC. Because \texttt{pymc3} takes a fully Bayesian approach, we actually obtain an estimate of the maximum a posteriori (MAP) solution, and the likelihood at this point can differ from the MLE when the priors are informative. To ensure that our model selection is not strongly impacted by this difference, we also consider the two Bayesian model selection methodologies provided by the \texttt{arviz} package: the Pareto-smoothed importance sampling leave-one-out statistic \citep[PSIS-LOO;][]{Vehtari17} and the Watanabe-Akaike Information Criterion, sometimes referred to as the Widely Applicable Information Criterion \citep[WAIC;][]{Watanabe10}. Both are methods for evaluating the out-of-sample predictive accuracy of a model, and more detailed comparisons between these statistics and the BIC can be found in e.g. \citet{Gelman14}. For each dataset, we verified that the detrending model with the lowest BIC also had the highest PSIS-LOO and WAIC values (within the uncertainties reported by \texttt{arviz}), indicating the highest predictive accuracy.

After selecting optimized detrending models for each night of data, we obtained final fits for each planet, in some cases jointly fitting multiple nights of data. In the final version of the fits we also included a jitter term on each night $\log(\sigma_\mathrm{extra})$, which quantifies the discrepancy between the photon noise and the true variance in the data (i.e. $\sigma^2$ = $\sigma_\mathrm{photon}^2$ + $\sigma_\mathrm{extra}^2$). Final $\sigma$ values for each light curve are provided in Table~\ref{table:log}, as well as the ratio $\sigma/\sigma_\mathrm{phot}$ between the achieved rms and the photon noise in the unbinned residuals. Allan deviation plots for each light curve are presented in Appendix~\ref{allan}. 

We summarize the resulting posteriors for our light-curve fits in Table~\ref{table:posteriors}. In this table, we also report the excess absorption at mid-transit, $\delta_\mathrm{mid}$. To derive this parameter, we first construct a comparison light curve at each step in the fit using the radius ratio measured in a nearby bandpass by another instrument along with our transit shape and limb darkening parameters. We then subtract the minimum value of the comparison light curve from the minimum value of our helium light curve at each step in the fit to obtain a distribution of $\delta_\mathrm{mid}$ values, which is summarized in Table~\ref{table:posteriors}. To construct the comparison light curves for WASP-69b, WASP-52b, HAT-P-18b, WASP-80b, and HAT-P-26b, we used the radius ratios from \textit{HST}/WFC3 G141 between 1110.8~nm and 1141.6~nm obtained by \citet{Tsiaras18}, i.e. the first point of the corresponding spectra in their Figure 4. We chose this wavelength bin because it was closest to 1083.3~nm. We also note that the water absorption features from the lower atmosphere are of order $\sim100$~ppm for these planets, far smaller than our precision on the helium transit depth, making the choice of baseline a negligible source of error in our analysis. 
WASP-177b and NGTS-5b are relatively recent discoveries and have not been observed by \textit{HST}; for the former planet, we fit jointly with existing \textit{TESS} (bandpass between 600~nm--1000~nm) observations of this system, and for the latter planet we use the reported radius ratio from NGTS \citep[an average across optical and NIR wavelengths;][]{Eigmuller19} since \textit{TESS} has not yet observed the system.

\begin{deluxetable*}{cccccccccc}[ht]
\tabletypesize{\scriptsize}
\tablecaption{Posteriors for the light-curve fits. \label{table:posteriors}}
\tablehead{\colhead{Planet} & \colhead{LDC} & \colhead{$P$ (days)}  & \colhead{$T_0$ (BJD $-$ 2450000)} & \colhead{$b$} & \colhead{$a/R_\star$} & \colhead{WIRC $R_\mathrm{p}/R_\star$} & \colhead{$u_1$} & \colhead{$u_2$} & \colhead{$\delta_\mathrm{mid}$ (\%)}}
\startdata
WASP-69b & free & $3.86813899_{-0.00000054}^{+0.00000053}$ & $7176.17789_{-0.00016}^{+0.00017}$ & $0.679_{-0.019}^{+0.019}$ & $11.64_{-0.28}^{+0.28}$& $0.1462_{-0.0025}^{+0.0026}$ & $0.47_{-0.30}^{+0.33}$ & $0.20_{-0.42}^{+0.40}$ & $0.506_{-0.066}^{+0.068}$ \\
& \textbf{fixed} & $3.86813940_{-0.00000052}^{+0.00000051}$ & $7176.17797_{-0.00016}^{+0.00016}$ & $0.691_{-0.017}^{+0.017}$ & $11.79_{-0.28}^{+0.27}$ & $0.1467_{-0.0017}^{+0.0017}$ & 0.38 & 0.12 & $0.512_{-0.048}^{+0.049}$ \\
\hline
WASP-52b & free & $1.749781181_{-0.000000099}^{+0.000000099}$ & $6770.05972_{-0.000040}^{+0.000040}$ & $0.596_{-0.018}^{+0.018}$ & $7.40_{-0.10}^{+0.10}$ & $0.1728_{-0.0039}^{+0.0038}$ & $0.26_{-0.19}^{+0.28}$ & $0.12_{-0.25}^{+0.31}$ & $0.28_{-0.14}^{+0.14}$ \\
& \textbf{fixed} & $1.749781183_{-0.000000099}^{+0.000000099}$ & $6770.059719_{-0.000040}^{+0.000039}$ & $0.594_{-0.016}^{+0.017}$ &  $7.373_{-0.096}^{+0.096}$ & $0.1729_{-0.0036}^{+0.0035}$ & 0.35 & 0.12 &  $0.29_{-0.13}^{+0.13}$ \\
\hline
HAT-P-18b & \textbf{free} & $5.50802974_{-0.00000077}^{+0.00000078}$ & $7276.256448_{-0.000098}^{+0.000099}$ & $0.365_{-0.049}^{+0.047}$ & $16.44_{-0.23}^{+0.23}$ & $0.1567_{-0.0047}^{+0.0046}$ & $0.67_{-0.32}^{+0.29}$ & $0.08_{-0.37}^{+0.41}$ & $0.70_{-0.16}^{+0.16}$ \\
& fixed & $5.50802973_{-0.00000077}^{+0.00000077}$ & $7276.256446_{-0.000097}^{+0.000099}$ & $0.412_{-0.041}^{+0.040}$ & $16.54_{-0.23}^{+0.23}$ & $0.1624_{-0.0036}^{+0.0035}$ & 0.38 & 0.12 & $0.84_{-0.13}^{+0.13}$ \\
\hline
WASP-80b & free & $3.06785259_{-0.00000018}^{+0.00000019}$ & $6671.49614_{-0.00004}^{+0.00004}$ & $0.216_{-0.022}^{+0.022}$ & $12.66_{-0.12}^{+0.12}$ &  $0.1708_{-0.0055}^{+0.0055}$ & $0.40_{-0.27}^{+0.36}$ & $0.21_{-0.40}^{+0.38}$ & $-0.01_{-0.22}^{+0.23}$ \\
& \textbf{fixed} & $3.06785259_{-0.00000018}^{+0.00000019}$ & $56671.496140_{-0.000040}^{+0.000040}$ & $0.219_{-0.022}^{+0.021}$ & $12.68_{-0.11}^{+0.11}$ &  $0.1716_{-0.0051}^{+0.0051}$ & 0.31 & 0.15 & $0.02_{-0.20}^{+0.20}$ \\
\hline
WASP-177b & free & $3.07172220_{-0.00000068}^{+0.00000068}$ & $7994.37141_{-0.00024}^{+0.00025}$ & $0.980_{-0.057}^{+0.071}$ & $9.51_{-0.30}^{+0.37}$ & $0.230_{-0.043}^{+0.056}$ & $0.62_{-0.45}^{+0.59}$ & $-0.03_{-0.42}^{+0.43}$ & $0.70_{-0.35}^{+0.35}$ \\
& \textbf{fixed} & $3.07172221_{-0.00000069}^{+0.00000068}$ & $7994.37142_{-0.00024}^{+0.00025}$ & $0.975_{-0.054}^{+0.069}$ & $9.52_{-0.31}^{+0.36}$ & $0.213_{-0.034}^{+0.050}$ & 0.36 & 0.12 & $0.53_{-0.28}^{+0.23}$ \\
\hline
HAT-P-26b & free & $4.23450038_{-0.00000067}^{+0.00000068}$ & $6892.590466_{-0.000097}^{+0.000100}$ & $0.288_{-0.111}^{+0.099}$ & $13.19_{-0.56}^{+0.51}$ & $0.0878_{-0.0058}^{+0.0055}$ & $0.26_{-0.19}^{+0.32}$ & $0.15_{-0.27}^{+0.36}$ & $0.30_{-0.11}^{+0.11}$ \\
& \textbf{fixed} & $4.23450039_{-0.00000066}^{+0.00000067}$ & $6892.590466_{-0.000099}^{+0.000099}$ &$0.290_{-0.110}^{+0.099}$ & $13.22_{-0.52}^{+0.48}$ & $0.0882_{-0.0054}^{+0.0051}$ & 0.35 & 0.12 &  $0.31_{-0.10}^{+0.10}$ \\
\hline
NGTS-5b & free & $3.3569892_{-0.0000024}^{+0.0000024}$ & $7740.35267_{-0.00026}^{+0.00026}$ & $0.675_{-0.046}^{+0.044}$ & $11.16_{-0.31}^{+0.31}$ & $0.188_{-0.014}^{+0.014}$ & $0.81_{-0.55}^{+0.57}$ & $-0.10_{-0.46}^{+0.52}$ & $1.07_{-0.50}^{+0.52}$\\
& \textbf{fixed} & $3.3569890_{-0.0000024}^{+0.0000024}$ & $7740.35267_{-0.00026}^{+0.00026}$ & $0.691_{-0.044}^{+0.041}$ & $11.20_{-0.31}^{+0.31}$ & $0.187_{-0.013}^{+0.013}$ & 0.36 & 0.12 & $1.02_{-0.46}^{+0.48}$
\enddata
\tablecomments{The ``LDC" column denotes the fitting strategy of leaving the limb darkening coefficients free or fixing them to a calculation from \texttt{ldtk} \citep{Husser13, Parviainen15}; the adopted strategy for each planet is noted in bold. The mid-transit excess depth $\delta_\mathrm{mid}$ is a derived parameter. Posteriors on the detrending weights are omitted for brevity.}
\end{deluxetable*}

\section{Results for Individual Planets}

\label{sec:lightcurves}
\subsection{WASP-69b}
WASP-69b is a Saturn-mass (0.26$M_\mathrm{J}$), Jupiter-sized (1.06$R_\mathrm{J}$) planet in a 3.87~day orbit around its K5 host star \citep{Anderson14}. Its low gravitational potential makes it an excellent target for atmospheric observations. To date, observations of this planet's lower atmosphere have revealed water absorption, sodium absorption, and a Rayleigh scattering slope indicating the presence of hazes \citep{Tsiaras18, Murgas20, Estrela21, Khalafinejad21}. Additionally, \citet{Nortmann18} detected helium escaping from the upper atmosphere of WASP-69b with two nights of CARMENES observations. We confirmed the strong absorption reported by these authors in our observation of this planet on UT 2019 Aug 16, which was initially published in \citet{Vissapragada20b}.

\begin{figure}[ht!]
\centering
\includegraphics[width=0.45\textwidth]{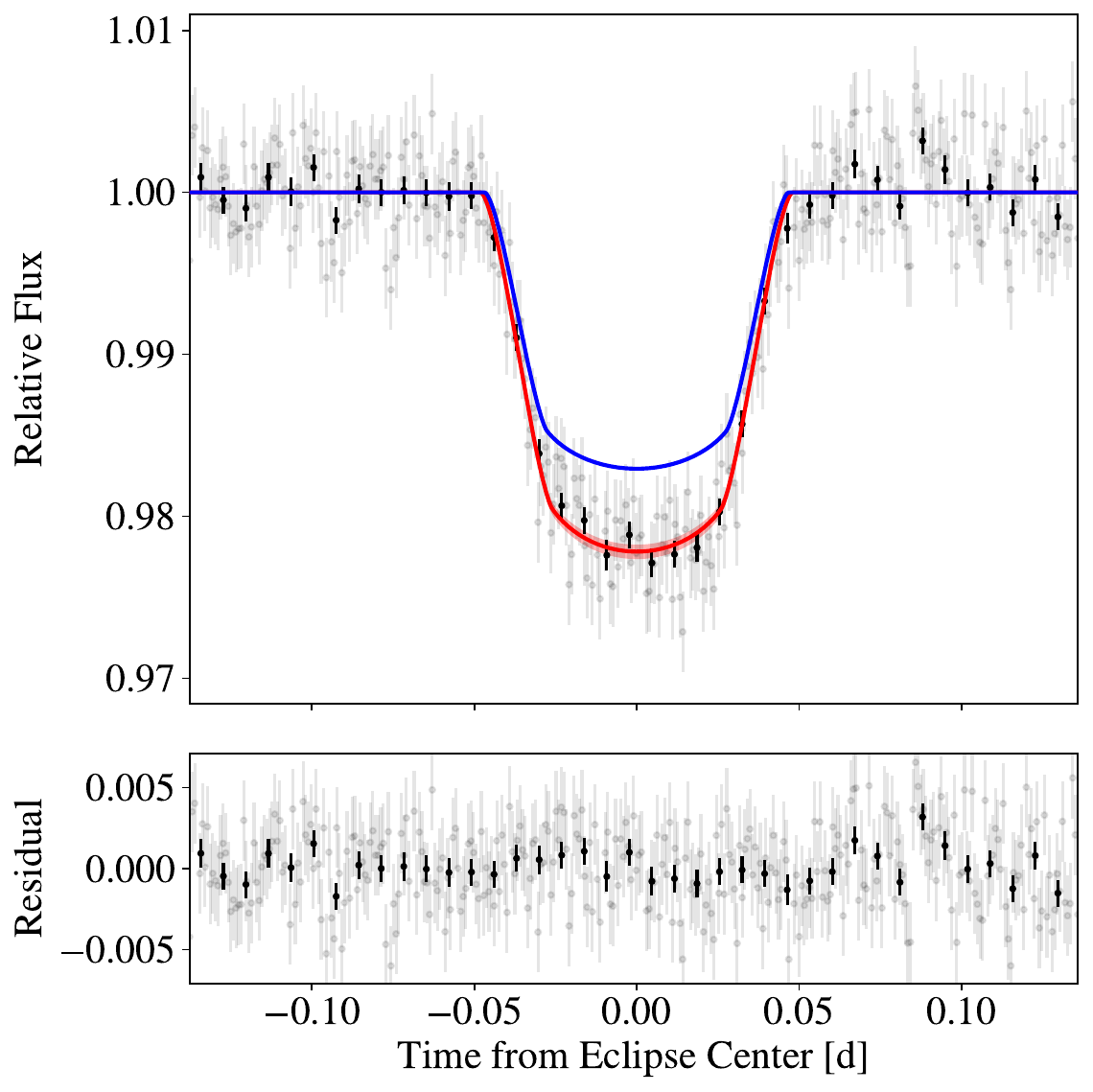}
\caption{Narrowband light curve (top) and residuals (bottom) for WASP-69b. Grey points are detrended data, which are binned to 10 minute cadence in black. The red curve indicates our best-fit model, with 1$\sigma$ uncertainty indicated by the red shading. The blue curve indicates the nominal light curve model (outside the helium bandpass).}
\label{fig:w69}
\end{figure}

Our transit modeling methodology has changed since that initial work, and we therefore re-fit the data to be consistent with the other planets in the sample. We found that the calculated limb darkening coefficients agreed well (within 2$\sigma$) with the joint distribution of retrieved quadratic limb darkening coefficients (Appendix~\ref{ldcs}), and the choice does not make much of a difference for the retrieved parameters. Therefore, we adopted the fixed limb darkening coefficient fit for this planet. The best-fit light curve, residual, and Allan deviation plot are shown in Figure~\ref{fig:w69}. We find the excess depth at mid-transit to be $0.512_{-0.048}^{+0.049}$\% in our bandpass, which is consistent with our previous work albeit with slightly larger uncertainty. In our previous work, the detrending model used the best-fit linear combination of comparison star vectors, effectively marginalizing over the detrending vector weights that we fit for in this work, so the previously-reported uncertainties were slightly underestimated.

\subsection{WASP-52b}
WASP-52b is a 0.46$M_\mathrm{J}$, 1.27$R_\mathrm{J}$ planet in a short 1.75~day orbit around its K2 host star \citep{Hebrard13}. Notably, the system appears to be relatively young: the discovery paper quotes a gyrochronological age of 400$^{+300}_{-200}$~Myr \citep{Hebrard13}, though this may be an underestimate \citep{Mancini17}. Its precarious position--just past the edge of the Neptune desert--was also noted by \citet{Owen18}. In the high-eccentricity migration framework, which is their preferred scenario for planets at the upper edge of the desert, this planet's position past the edge of the desert suggests that it should have been tidally disrupted. To resolve this discrepancy, \citet{Owen18} suggest that WASP-52b's radius may have been smaller during migration (with a correspondingly smaller tidal disruption radius), and that it underwent radius inflation only recently. Other published observations of WASP-52b indicate that this planet has a relatively flat transmission spectrum \citep{Kirk16, Chen17, Louden17, Mancini17, Alam18, May18}, although sodium, potassium, and H$\alpha$ absorption have all been detected at high spectral resolution \citep{Chen17, Alam18, Chen20}.

We first searched for helium in the atmosphere of this planet on UT 2019 Sep 17, and published the results of these observations in \citet{Vissapragada20b}. In this work, we re-fit that light curve jointly with a new observation taken on UT 2020 Aug 04. Observing conditions were worse on the second night, which rendered two faint comparison stars from the first night unobservable, but better positioning of the target star allowed us to add one extra comparison star that was not in the field of view on the first night. 

\begin{figure}[ht!]
\centering
\includegraphics[width=0.45\textwidth]{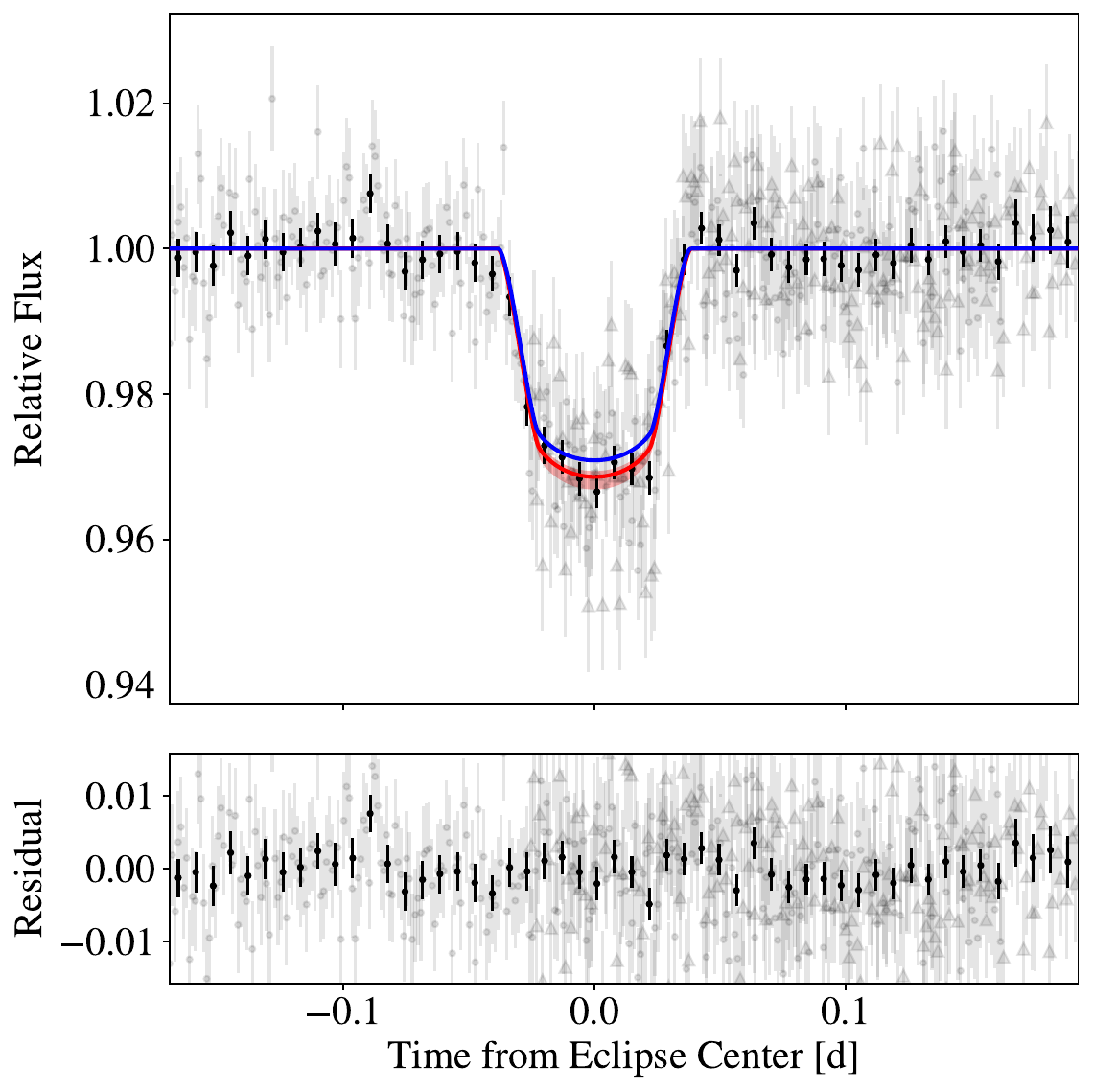}
\caption{Same as Figure~\ref{fig:w69}, but for WASP-52b. Points with circular markers indicate the first night of data collection and points with triangular markers indicate the second night.}
\label{fig:w52}
\end{figure}

For this planet, we adopted a fixed limb darkening coefficient fit, as we found that the calculated limb darkening coefficients agreed well with the joint distribution of retrieved quadratic limb darkening coefficients (Appendix~\ref{ldcs}), and the choice did not make a difference for the retrieved parameters. In separate retrievals, we constrained the excess absorption to be $0.22_{-0.13}^{+0.14}$\% on the first night, with a 95th-percentile upper limit of 0.44\%, and $1.02_{-0.41}^{+0.39}$\% on the second night with a 95th-percentile upper limit of 1.68\%. When fitting the datasets jointly, we found an excess absorption of $0.29_{-0.13}^{+0.13}\%$, corresponding to a tentative detection at $2.2\sigma$ confidence. The joint fit results are shown in Figure~\ref{fig:w52}.

\subsection{HAT-P-18b}
HAT-P-18b is a Jupiter-sized (1.00$R_\mathrm{J}$), low-mass (0.20$M_\mathrm{Jup}$) planet in a 5.5~day orbit around its K2 host star \citep{Hartman11a}. Despite its faintness ($J = 10.8$), the low gravitational potential of this planet makes it amenable to atmospheric characterization, and a Rayleigh scattering slope in HAT-P-18b's transmission spectrum was detected by \citet{Kirk17}. We previously detected a metastable helium signature for this planet in \citet{Paragas21}. The fitting method we employ in this survey is somewhat different from the one in \citet{Paragas21}, so we re-fit the observations here. Rather than fitting jointly with \textit{TESS} observations of the system as we did previously, we fit the WIRC data alone and compare to the precise transit depth from \textit{HST}/WFC3 G141 \citep{Tsiaras18}. We prefer this approach because the spectrophotometric bin we use from WFC3 (1110.8~nm-1141.6~nm) is narrower and closer to the helium filter bandpass than the \textit{TESS} bandpass (600~nm-1000~nm), and the depth in this bin is more precise than the \textit{TESS} depth ($\sim100$~ppm precision for WFC3 vs. $\sim500$~ppm for \textit{TESS}). Additionally, in the previous work we reported $(R_\mathrm{p}/R_\star)_\mathrm{WIRC}^2 - (R_\mathrm{p}/R_\star)_\mathrm{comparison}^2$ as the excess absorption, but this is not necessarily equal to the difference in transit depth between the WIRC light curve and comparison light curve, especially when there is strong stellar limb darkening in the helium line (as we infer for HAT-P-18b). To account for this, the comparison light curve should be constructed taking into account limb darkening in the helium bandpass, which we do when obtaining the excess absorption $\delta_\mathrm{mid}$ in this work (see Section~\ref{sec:modeling}).

\begin{figure}[ht!]
\centering
\includegraphics[width=0.45\textwidth]{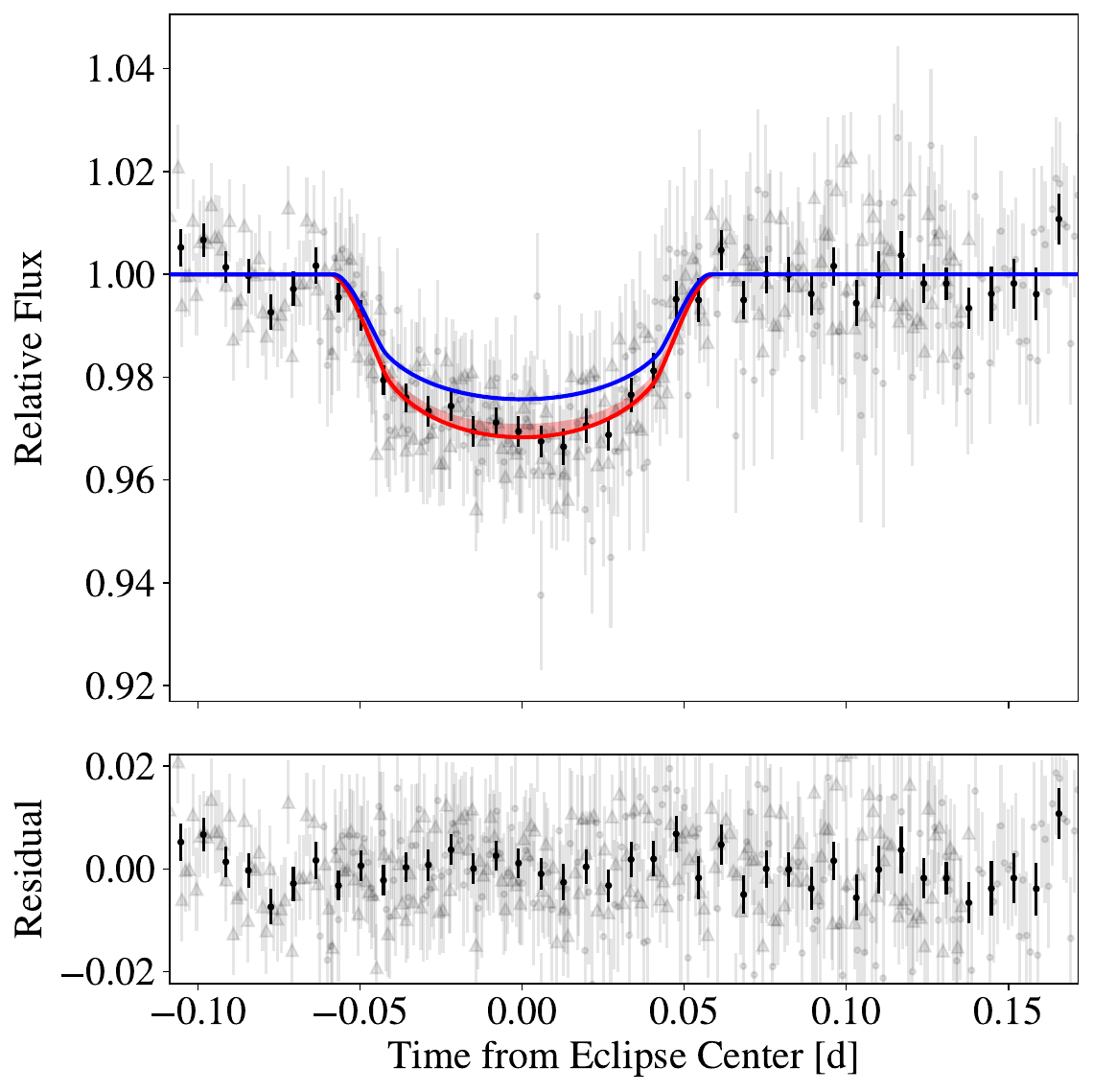}
\caption{Same as Figure~\ref{fig:w69}, but for HAT-P-18b. Points with circular markers indicate the first night of data collection and points with triangular markers indicate the second night.}
\label{fig:hp18}
\end{figure}

For HAT-P-18b, we adopted the free limb darkening coefficient fit. The fixed limb darkening coefficients disagreed with the retrieved coefficients by more than 2$\sigma$ in the joint posterior (Appendix~\ref{ldcs}), and the choice was consequential for the final fitted parameters, so we choose to adopt the free limb darkening coefficient solution. When fitting each dataset independently, we obtained $(R_\mathrm{p}/R_\star)^2$ values of $2.14_{-0.24}^{+0.23}$\% and $2.56_{-0.16}^{+0.17}$\% for the first and second nights, respectively. These are consistent to 1$\sigma$ with the values obtained by \citet{Paragas21}, who fit the WIRC data together with data from \textit{TESS} to obtain $(R_\mathrm{p}/R_\star)^2$ values of $2.11\pm0.25$\% for the first night and $2.35\pm0.14$\% for the second night. We then fit the WIRC datasets jointly and show the results in Figure~\ref{fig:hp18}. In the joint fit, we find $(R_\mathrm{p}/R_\star)^2 = 2.46\pm0.15$\%, which is about 1$\sigma$ larger than the result from \citet{Paragas21} of $2.29\pm0.12$\%. Our result for the excess depth is $0.70\pm0.16$\%, again slightly larger than our previously-reported signal of $0.46\pm0.12\%$.

\subsection{WASP-80b}
WASP-80b is a Jupiter-sized (1.00$R_\mathrm{J}$), 0.54$M_\mathrm{J}$ planet in a 3.07~day orbit around its K7/M0 host star \citep{Triaud13, Triaud15}. At low resolving power, the transmission spectrum of this planet is quite flat \citep{Fukui14, Mancini14, Turner17, Parviainen18, Kirk18}. Multiple groups running self-consistent wind-launching simulations have reported WASP-80b to be an excellent candidate for mass-loss observations \citep{Salz15, Salz16a, Caldiroli22}. However, \citet{Fossati22} recently obtained four nights of high-resolution data on WASP-80b and did not detect any metastable helium signature, setting an upper limit of 0.7\% absorption in the line core. 

We observed a full transit of WASP-80b on UT 2020 Jul 09. During the first half of observations the detector started reading large numbers of negative counts relatively infrequently until the issue became more persistent near ingress. The instrument was reset near the middle of the transit, which fixed the problem. Prior to analysis, frames in the first half with $>0.1\%$ of pixels reading negative counts were discarded. Additionally, the pre-reset data exhibited stronger systematics than the post-reset data, so we treated each of the two halves as separate observations (i.e., allowing each half of the light curve to have different coefficients for the systematics model) and fit them jointly. We optimized the aperture size using the second half of data only, finding an optimized radius of 8~px.

\begin{figure}[ht!]
\centering
\includegraphics[width=0.45\textwidth]{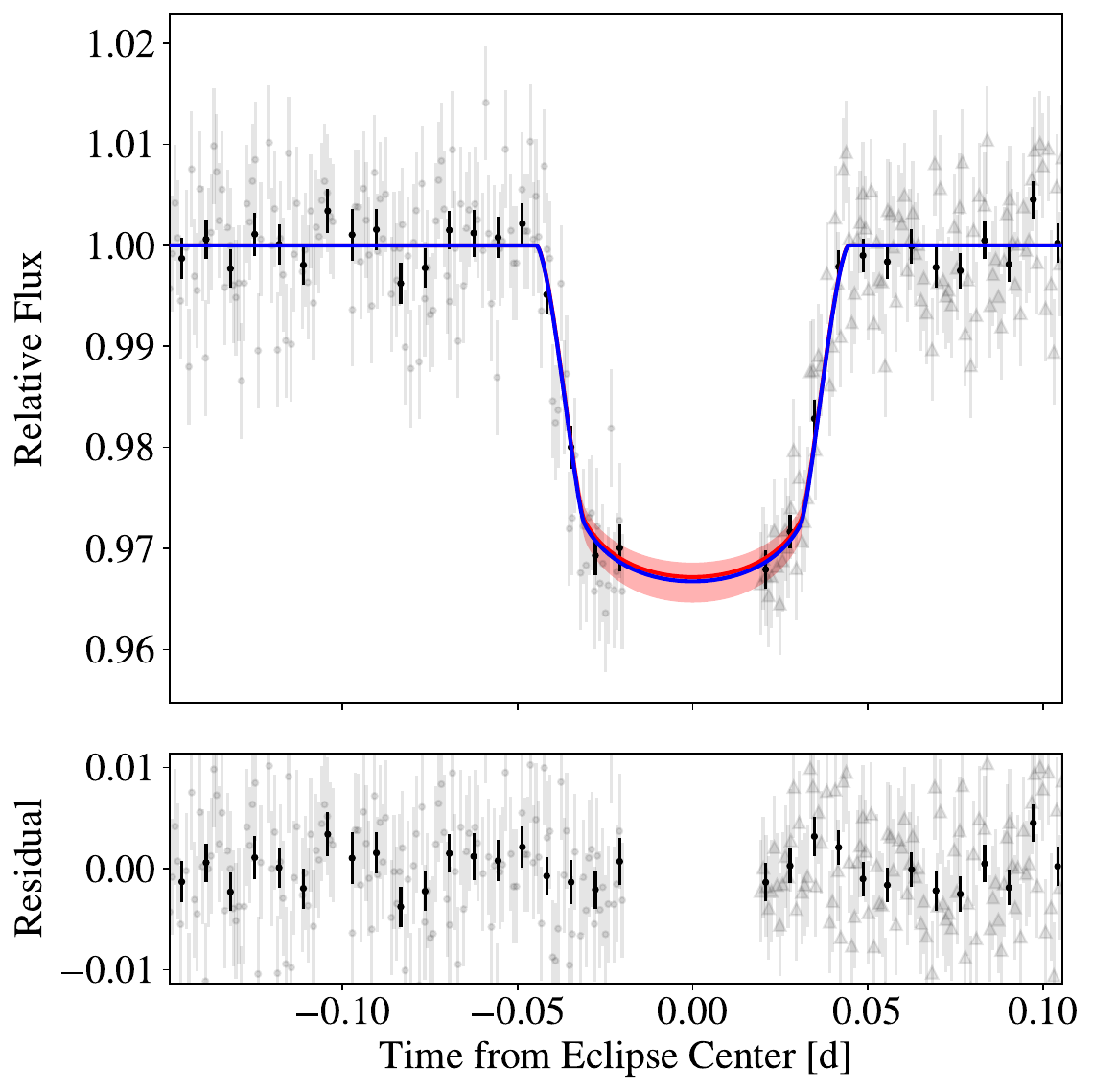}
\caption{Same as Figure~\ref{fig:w69}, but for WASP-80b. Points with circular markers indicate the first half of the observations (prior to the instrument reset) and points with triangular markers indicate the second half.}
\label{fig:w80}
\end{figure}

We adopted the fixed limb darkening coefficient fit for WASP-80b, as the calculated limb darkening coefficients agreed well with the joint distribution of retrieved quadratic limb darkening coefficients (Appendix~\ref{ldcs}), and the choice did not make a difference for the retrieved parameters. We did not detect any excess absorption in our data, finding $\delta_\mathrm{mid} = 0.02_{-0.20}^{+0.20}$ with a 95th-percentile upper limit of $0.35\%$ excess absorption in our bandpass. The fit results are shown in Figure~\ref{fig:w80}. Our non-detection agrees with the more sensitive observations from \citet{Fossati22}. 

\subsection{WASP-177b}
WASP-177b is a highly-inflated (1.6$R_\mathrm{J}$), 0.5$M_\mathrm{J}$ planet in a 3.07~day orbit around its K2 host star \citep{Turner19}. This planet has an exceptionally low density (similar to WASP-107b), making it a good target for transmission spectroscopy. However, the planetary transit is grazing, which makes it more difficult to extract precise constraints on the wavelength-dependent planet-star radius ratio. 

We observed a full transit of this planet on UT 2020 Oct 2. \textit{TESS} also observed this planet (TOI-4521.01) at 2~minute cadence in Sector 42 from UT 2021 Aug -- 2021 Sep. Given the grazing nature of the transit and the relatively small number of comparison light curves in the literature, we decided to fit our Palomar/WIRC light curve jointly with the TESS light curve. This allowed us to better capture covariances between the radius ratio, impact parameter, scaled semi-major axis, and limb darkening parameters while also giving us a reference transit depth to which we could compare the WIRC measurement. The \textit{TESS} portion of the fit was carried out similarly to our fit for HAT-P-18b in \citet{Paragas21}; the only additional parameters we included were separate \textit{TESS} limb darkening parameters, the radius ratio in the \textit{TESS} bandpass, and an error re-scaling term.

For a grazing geometry it is difficult to differentiate excess absorption associated with the planet from strong limb darkening, so we first tried a fit with free limb darkening coefficients. However, we found that the data were uninformative on the quadratic limb darkening coefficients (see Appendix~\ref{ldcs}), so we defaulted to using fixed limb darkening coefficients instead. The results of the joint fit are shown in Figure~\ref{fig:w177}. The fitted TESS lightcurve is given in Appendix~\ref{tessfit}. Using the \emph{TESS} transit depth as our baseline, we constrain the excess depth to be 0.53$^{+0.23}_{-0.28}$\%, a non-detection with a 95th-percentile upper limit of 0.90\%. 

\begin{figure}[ht!]
\centering
\includegraphics[width=0.47\textwidth]{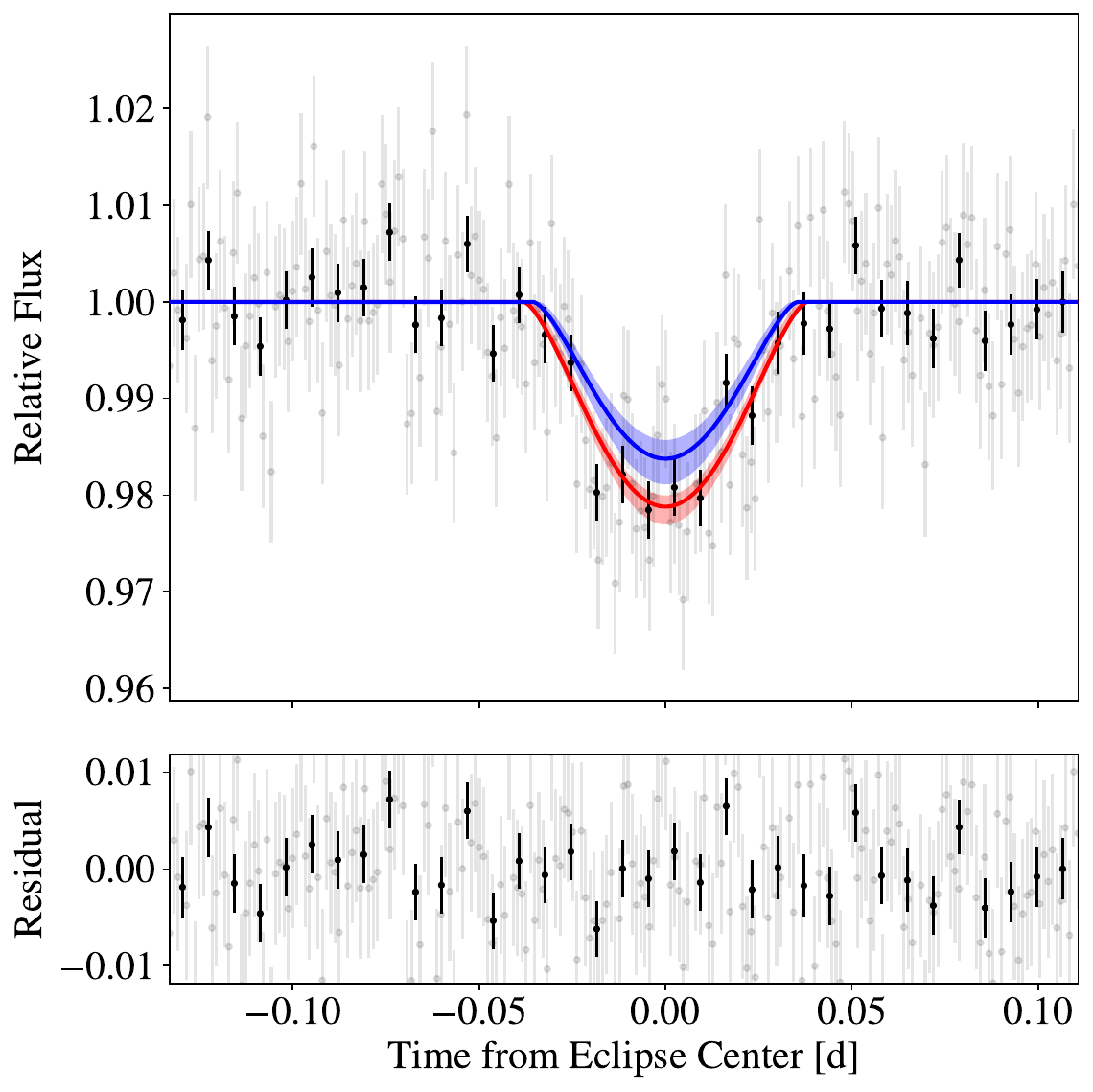}
\caption{Same as Figure~\ref{fig:w69}, but for WASP-177b. The blue shading indicates the non-negligible uncertainty on the baseline transit model for this planet.}
\label{fig:w177}
\end{figure}

\subsection{HAT-P-26b}
HAT-P-26b resides near the lower edge of the Neptune desert, and is the smallest planet in our sample. With a radius of a 0.57$R_\mathrm{J}$ and a mass of 0.059$M_\mathrm{J}$, this planet orbits its K1 host star every 4.23~days \citep{Hartman11b}. The Neptune-mass planet has a well-constrained heavy element abundance thanks to the presence of strong water bands in the red-optical and near-infrared \citep{Stevenson16, Wakeford17} as well as the potential presence of metal hydride features \citep{MacDonald19} and \textit{Spitzer} secondary eclipse depth ratios \citep{Wallack19}. At optical wavelengths, HAT-P-26b's alkali features appear to be obscured by a cloud layer \citep{Wakeford17, Panwar21}. 

This planet is an excellent target for observations of atmospheric escape. It is larger and less massive than HAT-P-11b \citep[a benchmark planet for metastable helium studies,][]{Allart18, Mansfield18}, and as a result has an inferred H/He envelope fraction approximately twice that of HAT-P-11b \citep{Lopez14}. The slightly earlier spectral type of HAT-P-26 (HAT-P-11 is a K4 dwarf) also puts it at the ``sweet spot'' for metastable helium observations as identified by \citet{Oklopcic19}. We observed this high-priority target three times: on UT 2021 Feb 18, 2021 May 1, and 2021 May 27. The weather on the first night was relatively poor, and the target was not optimally positioned. Between the first and second nights of observation, we shifted the placement of the target on the detector in order to include an additional comparison star, which greatly increased the final precision.

We do not detect the transit in the first night of data, for which we obtained an overall $R_\mathrm{p}/R_\star = 0.034_{-0.023}^{+0.028}$. As expected, this first night is much noisier than the latter two (Table~\ref{table:log}): the rms scatter in the unbinned residuals was 1.79\% (2.7$\times$ the photon noise) in the first observation, whereas it was 0.82\% (1.6$\times$ the photon noise) and 0.61\% (1.4$\times$ the photon noise) on the second and third nights, respectively. In order to avoid biasing the final joint fit, we therefore only included the latter two nights of photometry. 

\begin{figure}[ht!]
\centering
\includegraphics[width=0.45\textwidth]{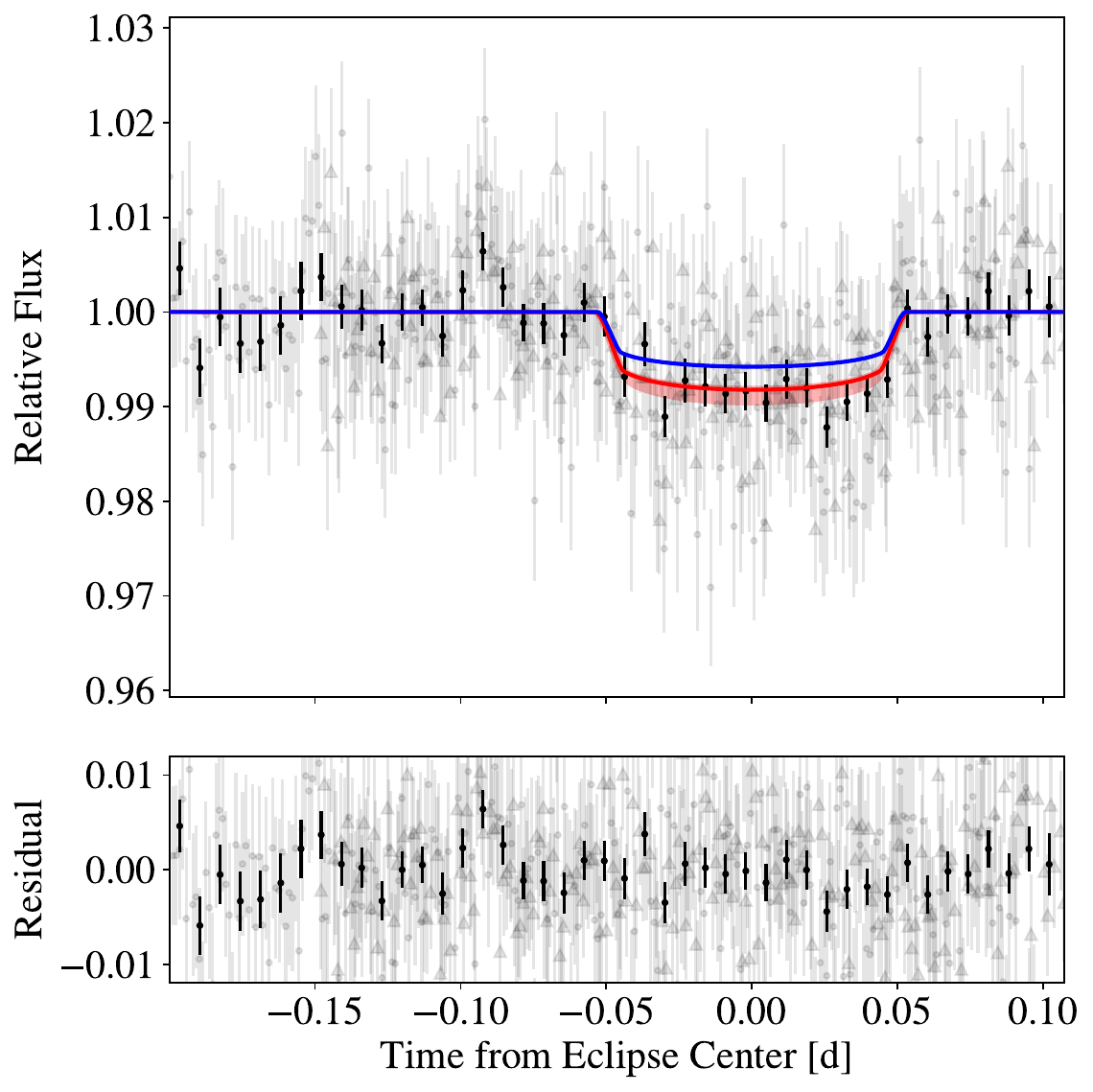}
\caption{Same as Figure~\ref{fig:w69}, but for HAT-P-26b. Points with circular markers indicate the second night of data collection and triangular markers indicate the third night. For both the second and third nights, the dashed red lines indicate 1.5$\times$ the photon noise.}
\label{fig:hp26}
\end{figure}

After optimizing the set of detrending vectors selected for each of the latter two nights, we obtained mid-transit excess depths of $0.43_{-0.17}^{+0.18}\%$ and $0.23_{-0.13}^{+0.13}\%$. We adopted the fixed limb darkening coefficient fit for HAT-P-26b, as the calculated limb darkening coefficients agreed well with the joint distribution of retrieved quadratic limb darkening coefficients (Appendix~\ref{ldcs}), and the choice did not make a significant difference for the retrieved parameters. The results of our joint fit to the latter two nights are shown in Figure~\ref{fig:hp26}. We obtained a mid-transit excess depth of $0.31_{-0.10}^{+0.10}$, evidence for helium absorption at 3.1$\sigma$ confidence.

\subsection{NGTS-5b}
NGTS-5b is a Jupiter-sized (1.1$R_\mathrm{J}$), sub-Saturn-mass (0.23$M_\mathrm{J}$) planet orbiting a K2 dwarf every 3.36~days \citep{Eigmuller19}. The planet is similar to WASP-69b in mass, radius, orbital period, and stellar host type, so despite the rather faint host star ($J = 12.1$) we identified it as a high-priority target. We observed two full transits of the planet: one on UT 2021 Apr 30 and one on UT 2021 May 27. 

This target had only one comparison star available in the field of view, and this along with the faintness of the target star created challenges for our standard systematics model approach. When testing different covariate combinations for the first night of data, the detrending models failed. With relatively poor weather conditions on that first night, the noisy comparison star light curve was severely underweighted in the systematics model when other covariates were included. When included in the joint fit, this underweighted model severely increased the magnitude of the correlated noise. We stress that unlike the first night for HAT-P-26b, which we discarded because there was no transit detected and the noise was nearly $3\times$ the photon noise limit, the transit is clearly detected in this dataset, and the issue stemmed purely from our approach to modeling the systematics. We therefore kept this dataset in the final model, but to avoid the sharp increase in correlated noise we did not use any additional covariates in the systematics model for the first night, and as such we do not report the model selection statistics for this night in Appendix~\ref{modelselection}. Conditions on the second night were more favorable, so the fit was more typical. 

\begin{figure}[ht!]
\centering
\includegraphics[width=0.45\textwidth]{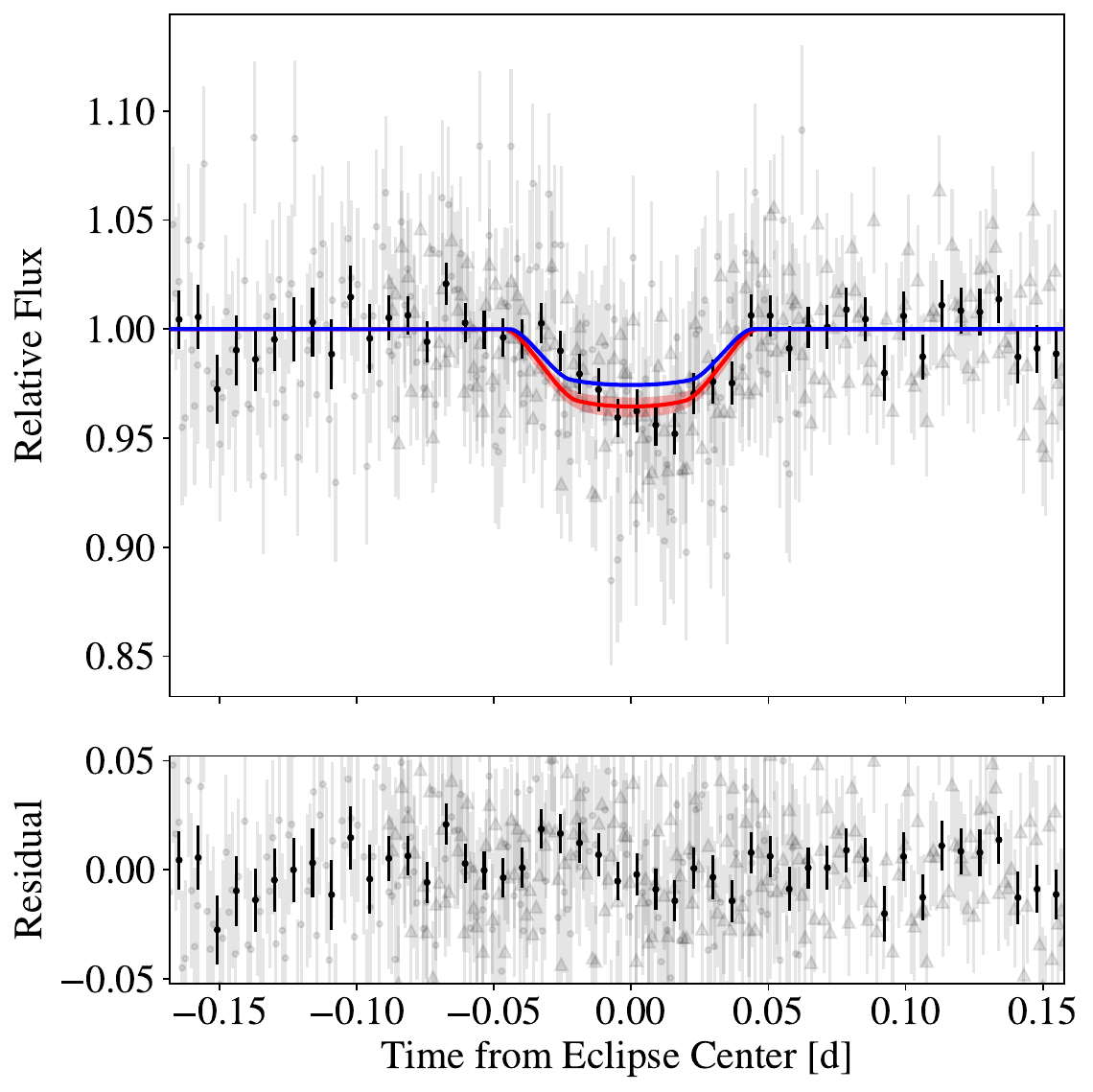}
\caption{Same as Figure~\ref{fig:w69}, but for NGTS-5b. Points with circular markers indicate the first night of data collection, and points with triangular markers indicate the second night.}
\label{fig:ngts5}
\end{figure}

We obtained mid-transit excess absorption values of $1.02_{-0.95}^{+0.99}$\% and $0.94_{-0.54}^{+0.55}$\% for the individual fits to the first and second nights, respectively. We proceeded to fit the datasets jointly, and the results are shown in Figure~\ref{fig:ngts5}. We adopted the fixed limb darkening coefficient fit for NGTS-5b as the calculated limb darkening coefficients agreed well with the joint distribution of retrieved quadratic limb darkening coefficients (Appendix~\ref{ldcs}) and the choice did not make a significant difference for the retrieved parameters. We obtained a final mid-transit excess absorption of $1.02_{-0.46}^{+0.48}$\%, a tentative detection at 2.2$\sigma$ confidence.

\section{Mass-Loss Modeling} \label{sec:massloss}
We can convert the excess depths from Section~\ref{sec:lightcurves} into constraints on planetary mass-loss rates by modeling the outflows with one-dimensional isothermal Parker winds \citep{Oklopcic18, Lampon20}. In this section, we use the \texttt{p-winds} code \citep{dosSantos22} to model our observations. Given a mass-loss rate assuming irradiation conditions at the substellar point $\dot{M}_\mathrm{substellar}$, a thermosphere temperature $T_0$, and a hydrogen fraction for the outflow $f_\mathrm{H}$, we first set the density and velocity profiles for the wind. Then, we use a high-energy stellar spectrum to calculate the ionization structure and level populations throughout the outflow. The level populations are used to compute the expected wavelength-dependent absorption signal at high resolving power, which we then convolve with our filter bandpass to compare to the measurements in Section~\ref{sec:lightcurves}.

The free parameters in this model are the mass-loss rate, thermosphere temperature, the hydrogen fraction, and the high-energy spectrum. We compute the model over a $50\times50$ grid of mass-loss rates $\dot{M}_\mathrm{substellar} = 10^{10}$~g~s$^{-1} - 10^{13}$~g~s$^{-1}$ and thermosphere temperatures $T_0 = 5000$~K$-15000$~K. The grid was uniformly spaced in $\log(\dot{M}_\mathrm{substellar})$ and $T_0$. We set the hydrogen fraction for all outflows to 0.9, corresponding to a 90-10 hydrogen-helium outflow. Finally, for the high-energy spectra we select proxy stars from the v2.2 panchromatic SEDs from the MUSCLES survey \citep[at 1~\AA~binning;][]{France16, Loyd16, Youngblood16}. We identify three K stars in MUSCLES that appear to be good proxies (i.e. they are the most similar in both spectral type and activity index $\log(R_\mathrm{HK}')$) for our targets: HD~85512, HD~40307, and $\epsilon$~Eridani. We model the late K stars ($T_\star \lesssim 4800$~K) in our sample using HD~85512's spectrum, and use HD~40307's spectrum for the early K stars. We use $\epsilon$~Eridani's spectrum to model the most active stars in our sample, including WASP-52 \citep[$\log(R_\mathrm{HK}') = -4.4$;][]{Hebrard13} and WASP-80 \citep[$\log(R_\mathrm{HK}') = -4.04$;][]{Fossati22}. We list the spectrum used for each star in Table~\ref{table:massloss}. We then scale each MUSCLES spectrum by the stellar radius and planetary orbital distance to obtain the high-energy spectrum incident at the top of the planet's atmosphere. We integrate each spectrum up to the hydrogen photoionization threshold 91.2~nm to obtain an estimate of the top-of-atmosphere XUV flux $F_\mathrm{XUV}$ for each planet, which are included in Table~\ref{table:massloss}.

\begin{deluxetable*}{cccccc}[ht]
\tablecaption{Mass-loss modeling summary. \label{table:massloss}}
\tablehead{\colhead{Planet}  & \colhead{MUSCLES spectrum}  & \colhead{$F_\mathrm{XUV}$} & \colhead{$\dot{M}$} & $\varepsilon$ & $M_\mathrm{env}/\dot{M}$\\
& & $10^3$~erg~cm$^{-2}$~s$^{-1}$ & $10^{10}~\mathrm{g~s^{-1}}$ & & Gyr}
\startdata
WASP-69b & HD~85512 & 2.3 &$6.3_{-4.9}^{+9.6}$ & $0.43_{-0.33}^{+0.66}$ & $250_{-150}^{+830}$\\
WASP-52b & $\epsilon$~Eridani & 36.4 &$2.7_{-2.2}^{+9.0}$ & $0.0079_{-0.0065}^{+0.0261}$ & $1010_{-780}^{+4710}$ \\
HAT-P-18b & HD~85512 & 1.1 & $3.7_{-2.7}^{+5.6}$ & $0.54_{-0.39}^{+0.81}$ & $320_{-190}^{+820}$ \\
WASP-80b & $\epsilon$~Eridani & 12.3 &$<1.4$ & $<0.046$ & $>12000$ \\
WASP-177b & HD~40307 & 8.9 & $<45$ & $<0.37$ & $>1300$ \\
HAT-P-26b & HD~40307 & 4.5 & $9.2_{-8.0}^{+25.0}$ & $0.52_{-0.45}^{+1.42}$ & $12.2_{-8.9}^{+77.6}$ \\
NGTS-5b & HD~40307 & 6.4 & $10.2_{-9.7}^{+33.8}$ & $0.16_{-0.15}^{+0.52}$ & $140_{-100}^{+2770}$
\enddata
\tablecomments{$F_\mathrm{XUV}$ corresponds to the XUV flux at the planet. We quote the 95\% confidence intervals for $\dot{M}$, $\varepsilon$, and $M_\mathrm{env}/\dot{M}$ as the marginalized $\dot{M}$ distributions are highly non-Gaussian (see Figure~\ref{fig:caldiroliplot}). Upper limits correspond to 95\% on the corresponding distributions.}
\end{deluxetable*}

After running an initial grid of models, we noticed that many of our winds had sonic points located outside the Roche radius. As discussed in \citet{MurrayClay09}, this happens when the tidal gravity term in the momentum equation for the wind is not taken into account. We show in Appendix~\ref{tidal} that the tidal gravity term appreciably alters the velocity and density profiles for the planets in our sample, and we update the \texttt{p-winds} code (in version 1.3) to account for the effects of the stellar gravity. We then re-calculate our model grids for each planet and use these grids to determine the range of mass-loss rates and thermosphere temperatures that match our observations. Next, we assess which combinations of $\dot{M}_\mathrm{substellar}$ and $T_0$ are energetically self-consistent using the methodology outlined in \citet{Vissapragada22}. If photoionization is the only heat source, certain combinations do not satisfy energy balance in the isothermal Parker wind, and we omit them from our final results. 

Finally, we divide the substellar-point mass-loss rates by a factor of 4 to obtain estimates for the overall mass-loss rate $\dot{M}$. This correction factor accounts for the fact that the planet is only irradiated over $\pi$ steradians rather than the $4\pi$ steradians assumed by our 1D models, so the true mass-loss rates will be about a factor of 4 smaller \citep{Salz16a, Vissapragada22}. Other prescriptions to correct for 2D effects exist \citep[e.g.][]{Odert20}, but they all agree well for outflows from low-gravity planets (like the ones in our sample) that are dominated by adiabatic cooling \citep{Caldiroli21}. Performing the radiative transfer using the 1D model rather than a 2D or 3D model also introduces error, but comparisons between 1D and 3D models suggest that this is a minor error term (relative to the uncertainty from the $\dot{M}-T_0$ degeneracy) except in the presence of strong stellar winds \citep{MacLeod22, Wang21a, Wang21b}.

\begin{figure*}[ht!]
\centering
\includegraphics[width=\textwidth]{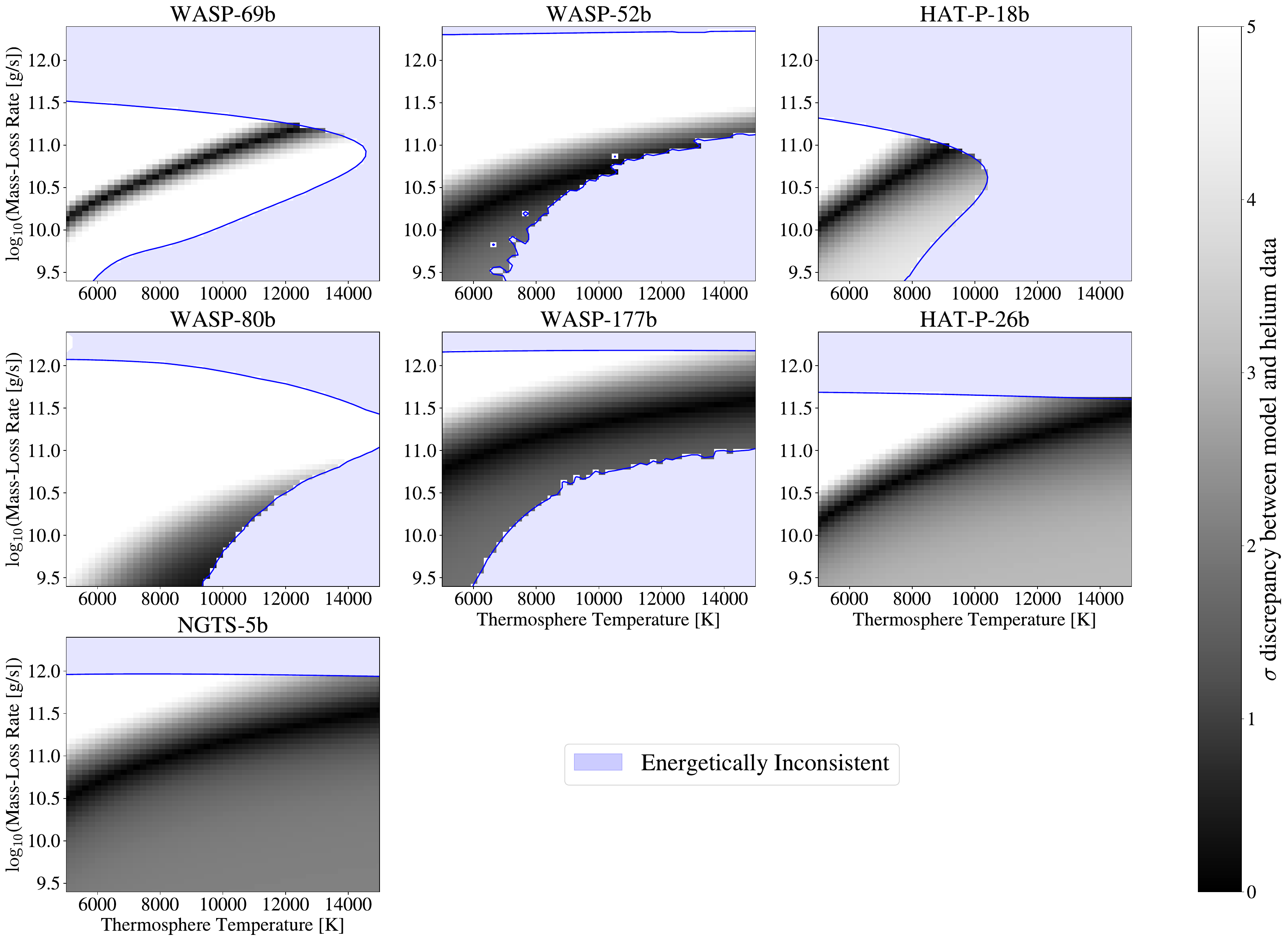}
\caption{Mass-loss models for the seven planets in our sample. The shading gives the $\sigma$ agreement between the observations and a 1D transonic Parker wind model characterized by the mass-loss rate $\dot{M}$ and thermosphere temperature $T_0$. Models that violate energy balance \citep{Vissapragada22} are rejected (blue shaded regions).}
\label{fig:massloss}
\end{figure*}

We show the resulting mass-loss constraints for each planet in Figure~\ref{fig:massloss}. We used a rejection sampling scheme with this grid of results to generate a list of 100,000 samples for each planet corresponding to the underlying joint distribution of mass-loss rates and thermosphere temperatures. We began by uniformly sampling temperatures from $T_0 \sim \mathcal{U}(5000, 150000)$ and $\log(\dot{M}_\mathrm{substellar}) \sim \mathcal{U}(10, 13)$. For each sample, we found the closest point on the grid in Figure~\ref{fig:massloss} with corresponding $x\sigma$ discrepancy between the model and the data. We accepted the sample only if $y > \mathrm{erf}(x/\sqrt{2})$, where $y$ is a random draw from $\mathcal{U}(0,1)$. 

With the final list of 100,000 samples in hand for each planet, we marginalize over the thermosphere temperature to obtain distributions for the mass-loss rates, which are summarized in Table~\ref{table:massloss}. We evaluate the significance of these mass-loss rates for the expected atmospheric lifetime of each planet by calculating the envelope loss timescale $M_\mathrm{env}/\dot{M}$. For HAT-P-26b we adopt an envelope mass fraction of 31.7\% \citep{Lopez14}. For the other higher-mass gas giants in the sample, we take $M_\mathrm{env}\sim M$. We give the resulting envelope loss timescales in Table~\ref{table:massloss}.

\section{Discussion} \label{sec:disc}
\subsection{Benchmarking Mass-Loss Models} \label{sec:benchmark}
Self-consistent, one-dimensional mass-loss models like the Pluto-CLOUDY Interface \citep[TPCI;][]{Salz15, Salz16a} and the ATmospheric EScape code \citep[ATES;][]{Caldiroli21,Caldiroli22} are crucial for interpretation of mass-loss-related phenomena in the exoplanet population, but to date there have been relatively few attempts to benchmark their predictions against measurements of present-day mass-loss rates for transiting planets. Here, we compare the inferred mass-loss rates from our seven-planet sample to predictions from the ATES code as a function of the incident XUV flux. Because \citet{Caldiroli22} provide analytical approximations for the efficiency as a function of incident XUV flux and gravitational potential, comparisons between the model and the data are simple and do not require expensive computations for each planet.

In order to place all seven planets on the same plot, we need to convert our measured mass-loss rates and XUV fluxes into scaled units that correct for planet-to-planet variations in gravitational potential and Roche radius. Scaled units can be obtained by starting from the energy-limited approximation from the planetary mass-loss rate \citep[e.g.][]{Caldiroli22}:
\begin{equation}
    \dot{M} = \frac{\varepsilon \pi R_\mathrm{p}^3F_\mathrm{XUV}}{KGM_\mathrm{p}},
    \label{energylim}
\end{equation}
where $\varepsilon$ is the mass-loss efficiency, $F_\mathrm{XUV}$ is the incident high-energy flux, and $K$ is the \citet{Erkaev07} Roche-lobe correction factor. Then, we can rewrite the expression in terms of the planetary density $\rho_\mathrm{p}$:
\begin{equation}
    K\dot{M} = \frac{3\varepsilon}{4G}\frac{F_\mathrm{XUV}}{\rho_\mathrm{p}}. \label{cald}
\end{equation}
Planets in the energy-limited regime should therefore exhibit a linear relation between the $K$-corrected mass-loss rate and the ratio $F_\mathrm{XUV}/\rho_\mathrm{p}$ \citep{Caldiroli22}.

As the incident XUV flux increases, however, self-consistent models predict that the outflow begins to lose substantial energy to radiative cooling, leading to a sub-linear dependence on $F_\mathrm{XUV}/\rho_\mathrm{p}$ beyond a threshold of $F_\mathrm{XUV}/\rho_\mathrm{p}\sim10^{4}$~erg~cm~g$^{-1}$s$^{-1}$ \citep{MurrayClay09, Salz16a, Salz16b, Caldiroli22}. Higher-gravity planets are also predicted to have low overall outflow efficiencies, as the atmospheres are more tightly bound and tend to re-emit more energy locally \citep{Owen16}. To an order of magnitude, this happens when the energy liberated by a single photoionization event ($\Delta E \sim10$~eV $\sim10^{-11}$~erg) exceeds the particle binding energy, which for a hydrogen atom ($m_\mathrm{H}\sim10^{-24}$~g) occurs roughly at a gravitational potential of $\phi_\mathrm{p}\sim\Delta E/m_\mathrm{H} \sim 10^{13}$~erg/g \citep[for a more thorough explanation, see][]{Caldiroli22}. More sophisticated numerical experiments place this threshold closer to $\phi_\mathrm{p} \gtrsim 10^{13.2}$~erg/g \citep{Salz16a, Caldiroli22}.

\begin{figure*}[ht!]
\centering
\includegraphics[width=\textwidth]{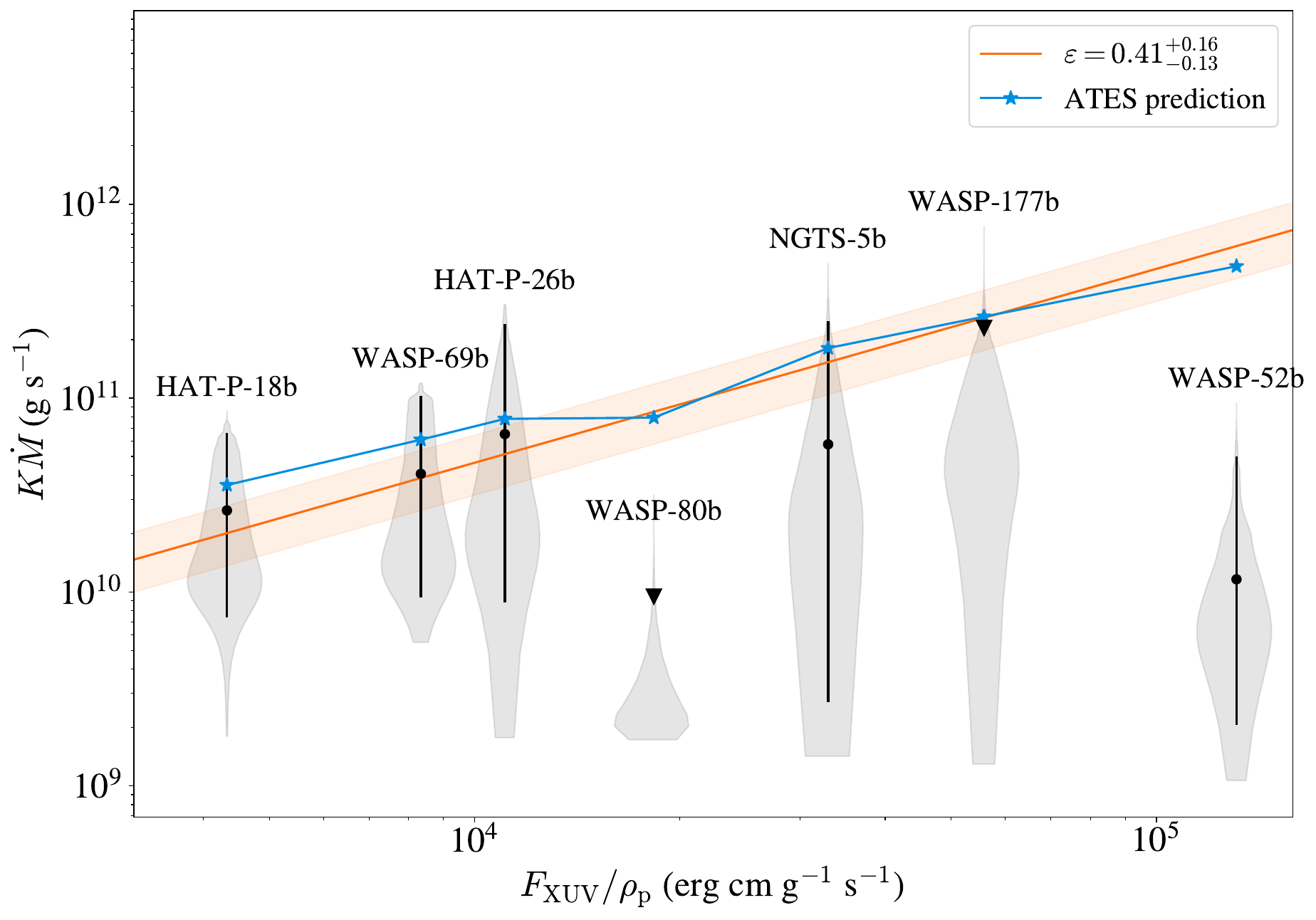}
\caption{$K$-corrected mass-loss rate $\dot{M}$ as a function of $F_\mathrm{XUV}/\rho_\mathrm{p}$ for all the planets in our sample, following \citet{Caldiroli22}. The shaded violins indicate the marginalized distribution of $K$-corrected mass-loss rates with the median and 95\% confidence interval for the distribution given by the black points and error bars, respectively. The triangle denotes the 95\% upper limit for WASP-80. The orange curve indicates the average inferred mass-loss efficiency $\varepsilon$, with the 1$\sigma$ uncertainty given by the shaded orange region. The blue curve indicates predictions from the ATES code \citep{Caldiroli21,Caldiroli22}.}
\label{fig:caldiroliplot}
\end{figure*}

In Figure~\ref{fig:caldiroliplot} we plot our marginalized and $K$-corrected $\dot{M}$ distributions as a function of $F_\mathrm{XUV}/\rho_\mathrm{p}$, along with the corresponding ATES model prediction for each planet. Immediately, two distinct populations of planets emerge in this parameter space. The first, comprised of HAT-P-18b, WASP-69b, HAT-P-26b, NGTS-5b, and WASP-177b, appear to agree quite well with the ATES predictions. The second group of planets, which includes WASP-80b and WASP-52b, appear to have mass-loss rates that are more than an order of magnitude lower than the model predictions. We discuss possible explanations for the surprisingly low inferred mass-loss rates of the second group in Section~\ref{sec:outliers}.

We use a bootstrap averaging method to estimate the mean mass-loss efficiency of planets in the first group. We omit WASP-52b and WASP-80b from this exercise because their helium signals appear to be affected by factors that are not included in our model, such as magnetic fields or stellar winds, as discussed in the next section. For the other planets, we first converted our marginalized $\dot{M}$ distribution for each planet into a distribution on $\varepsilon$ using Equation~(\ref{cald}), and these distributions are summarized in Table~\ref{table:massloss}. We then took a random draw from each planet's $\varepsilon$ distribution and averaged the draws to get an estimate for the mean efficiency. We repeated this procedure 100,000 times to obtain a final estimate of $\varepsilon = 0.41_{-0.13}^{+0.16}$. We overplot this average $\varepsilon$ model in Figure~\ref{fig:caldiroliplot}. 

The measured mass loss constraints for all five planets in the first group overlap with the average-$\varepsilon$ model at 2$\sigma$ confidence, and the model is a reasonable approximation to the ATES \citep{Caldiroli22} predictions for our sample as well. ATES also predicts that the mass-loss efficiency should level off with increasing $F_\mathrm{XUV}/\rho_\mathrm{p}$ as compared to a fixed-$\varepsilon$ curve. Unfortunately, the uncertainties in the inferred mass-loss rates for our sample, which are dominated by the degeneracy between mass-loss rate and thermosphere temperature, are too large to differentiate between this model and a constant-$\varepsilon$ model. Resolving the line profiles of these planets with high-resolution transmission spectroscopy can help break the degeneracy \citep[e.g.][]{dosSantos22}.

We can compare our measured efficiency to other giant planets ($R_\mathrm{p} > 4R_\Earth$) orbiting K stars with reported helium detections in the literature. The reported absorption for WASP-107b \citep{Spake18, Allart19, Kirk20, Spake21} is well-matched by the three-dimensional hydrodynamical models of \citet{Wang21b}, who report a mass-loss rate of approximately $2\times10^{11}$~g~s$^{-1}$. Using the scaled MUSCLES spectrum of the similarly active star $\epsilon$~Eridani \citep[$\log(R_\mathrm{HK}') = -4.44$;][]{Noyes84} to calculate the XUV flux for WASP-107b \citep[$\log(R_\mathrm{HK}') = -4.43$; calculated using the S-value from][]{Spake18}, we obtained an outflow efficiency of 0.34, in good agreement with our measured efficiency. The signal observed for HAT-P-11b \citep{Allart18, Mansfield18} has been modeled using \texttt{p-winds} using the scaled MUSCLES spectrum of HD 40307 as a proxy for the stellar XUV spectrum, and \citet{dosSantos22} report a 1D mass-loss rate of $2.3_{-0.5}^{+0.7}\times10^{10}$~g~s$^{-1}$ (with stellar limb darkening taken into account). Dividing this result by 4 to get the overall mass-loss rate (for consistency with the method outlined in Section~\ref{sec:massloss}), we obtain an outflow efficiency of ${0.170}_{-0.037}^{+0.052}$, somewhat smaller than our measured efficiency but consistent within 2$\sigma$. 

Finally, the signal observed for HD 189733b \citep{Salz18, Guilluy20} was modeled using a Parker wind methodology by \citet{Lampon21a}. For their 90/10 H/He composition model, they found a mass-loss rate at $T = 12000$~K of about $4\times10^{9}$~g~s$^{-1}$, with significant spread due to the $\dot{M}-T_0$ degeneracy. Again dividing by 4 to account for the multidimensional outflow, this corresponds to an efficiency of 0.0023, much smaller than our measured efficiency. A small efficiency is expected: the relatively large mass of HD 189733b puts it in a regime where radiative losses dominate the cooling budget rather than the adiabatic expansion of the atmosphere \citep{Salz16a, Caldiroli21, Caldiroli22}, so the outflow is more similar to WASP-52b and WASP-80b than the other planets in our sample. Similarly to WASP-52b and WASP-80b, the planet has a smaller semimajor axis \citep[$a = 0.031$;][]{Bouchy05} and more active host star \citep[$\log{(R_\mathrm{HK}')}= -4.501$;][]{Knutson10} than the other planets in our sample. The inferred mass-loss rate for HD 189733b at a 90/10 composition is still at least a factor of a few smaller than predicted by 1D hydrodynamical models, which give mass-loss rates of about $10^{9.5-10.0}$~g~s$^{-1}$ for this planet \citep{Salz16a, Caldiroli21, Caldiroli22}. \citet{Lampon21a} resolve this discrepancy by invoking a hydrogen-rich composition for the outflow, but this is not the only factor that can achieve a reduction of the helium signal. In the next section, we consider a range of explanations for the smaller-than-expected helium signals on WASP-52b and WASP-80b that could apply to HD 189733b as well. 

\subsection{Low Mass-Loss Rates for WASP-80b and WASP-52b}
\label{sec:outliers}
We now consider potential explanations for WASP-80b and WASP-52b's low apparent mass-loss rates, which are discrepant with both the ATES model predictions and the rest of the planets in our sample. We note that these two planets have the smallest semimajor axes and the largest $\log(R'_\mathrm{HK})$ values (Table~\ref{table:sample}) in our sample; this may or may not have anything to do with their low observed mass-loss rates. In this section, we consider five possible explanations: overestimated XUV fluxes, the XUV spectral shapes of their host stars, stronger-than-predicted stellar winds, outflow composition, and magnetic fields.

If the XUV flux of the host star was overestimated, it would cause us to overpredict the magnitude of a planet's mass loss rate. To test this effect, we decreased the incident XUV flux by a factor of ten in the \texttt{p-winds} models for WASP-52b and WASP-80b and recomputed the mass-loss efficiencies. Averaging the two efficiency parameters using the same bootstrapping method from Section~\ref{sec:benchmark}, we found a mean efficiency of $0.130_{-0.048}^{+0.064}$. This is still smaller than our inferred efficiency for the other five planets, but the distributions overlap at the $2\sigma$ level. We conclude that the XUV fluxes would have to be overestimated by at least a factor of ten for the mass-loss efficiencies to come into reasonable agreement with the rest of the sample.

For WASP-80, \citet{Fossati22} combined archival ROSAT and XMM-Newton observations with a new Gaia distance to calculate a precise X-ray flux measurement of $L_\mathrm{X} = 4.85^{+0.12}_{-0.23}\times10^{27}$~erg/s. They used the \citet{King18} relations to scale this value into the UV, obtaining a final $F_\mathrm{XUV} \approx 6300$~erg/cm$^2$/s. Two other papers independently derive this star's X-ray flux and find values that are approximately a factor of three smaller \citep{Monsch19} and a factor of two larger \citep[][who obtain a flux close to our adopted value in Table~\ref{table:massloss}]{Foster22} than that obtained by \citet{Fossati22}. Despite these methodological differences, factor-of-a-few variations in the XUV flux are not enough to reconcile the tight upper limit on this planet's helium absorption signal with model predictions. For WASP-52, which was observed by Chandra (ACIS), \citet{Monsch19} found $L_\mathrm{X} = 3.1^{+1.2}_{-1.0}\times10^{28}$~erg/s. Using the X-ray to EUV scaling relation for Chandra (ACIS) from \citet{King18}, we obtain an XUV irradiance of $F_\mathrm{XUV} \approx 38,000$~erg/cm$^2$/s. This is quite close to our assumed value (see Table~\ref{table:massloss}). We therefore conclude that, just as with WASP-80, differences in assumed XUV flux are unlikely to explain this planet's low helium absorption signal.

While variations in the integrated XUV flux are unlikely to explain the helium observations for WASP-80b and WASP-52b, the assumed spectral shape may play a role. In order to create a significant population of metastable helium, there must be enough EUV photons to ionize ground-state helium in the outflow, which then recombines into the metastable state \citep{Oklopcic18, Oklopcic19}. As demonstrated by \citet{Poppenhaeger22}, many of the stellar EUV photons that ultimately ionize helium come from coronal iron lines. Iron has a relatively low first ionization potential (FIP), and in very active, low-mass stars, the abundance of species with low FIPs often appear to be diminished \citep[termed the inverse first ionization potential effect, or iFIP;][]{Brinkman01, Gudel01, Wood18}. WASP-52 and WASP-80 are by far the most active stars in our sample, so they may be affected by this phenomenon. Our assumed proxy star for these two planets, $\epsilon$ Eridani, does not exhibit the iFIP effect and has strong coronal iron lines \citep{Poppenhaeger22}; this might cause us to overestimate their metastable helium populations in our models.

It is also worth noting that the models we are using in this study assume spherical symmetry, but not all outflows are spherically symmetric. If the outflow is confined into a comet-like shape by a strong stellar wind, it could reduce the magnitude of the helium absorption \citep[e.g.,][]{McCann19, Carolan20, MacLeod22}. This phenomenon has been directly detected for the hot Jupiter WASP-107b \citep{Khodachenko21, Spake21, Wang21b}, and may also explain the relatively weak helium signal measured for the mini-Neptune TOI-560.01 \citep{Zhang22}. Importantly, both WASP-52 and WASP-80 appear to be relatively young, suggesting that they may have correspondingly enhanced stellar winds. WASP-52 \citep[$v\sin{i}\approx$3.6~km/s with a photometric rotation period of 16.4~days,][]{Hebrard13} and WASP-80 \citep[$v\sin{i}\approx$3.55~km/s,][]{Triaud13} both rotate rapidly, with estimated gyrochronological ages of less than 1 Gyr \citep{Barnes07, Hebrard13, Triaud13}. They are also the two most active stars in our sample, with $B-V$ colors and $\log{R_\mathrm{HK}'}$ values that also suggest ages as young as a few hundred Myr \citep{Mamajek08}. However, some caution is warranted as stellar angular momenta are known to be affected by giant planets on close in-orbits \citep[e.g.][]{Lanza10, Poppenhaeger14, Mancini17}.

\citet{Fossati22} modeled the impact of stellar wind on WASP-80b using a three-dimensional hydrodynamical model \citep{Shaikhislamov21, Khodachenko21}, and showed that it does not suppress the helium signal enough to match their non-detection. They modeled stellar mass-loss rates up to $10^{13}$~g~s$^{-1}\sim8\dot{M}_\Sun$, where the solar mass-loss rate is $\dot{M}_\Sun\approx2\times10^{-14}M_\Sun$~yr$^{-1}$ \citep[e.g.][]{Wood05}, but at early times the wind can be even stronger. \citet{Wood05} used astrospheric Lyman-$\alpha$ absorption to infer mass-loss rates of 30$\dot{M}_\Sun$ and 100$\dot{M}_\Sun$ for $\epsilon$~Eridani and 70 Oph, both relatively young K dwarf stars. We can use the aforementioned X-ray fluxes along with the \citet{Wood05} relation to estimate the stellar mass-loss rate for WASP-80. We obtain a mass-loss rate of $\sim4\dot{M}_\Sun$, well within the range modeled by \citet{Fossati22}. Using the same relation, we obtain a mass-loss rate of $\sim35\dot{M}_\Sun$ for WASP-52. We conclude that a strong stellar wind might suppress the helium signal for WASP-52b, but it is unlikely to explain WASP-80b.

The composition of the outflow could also affect the magnitude of the observed signal. It has been suggested \citep[e.g.][]{Lampon20, Lampon21a} that outflows from giant planets may be depleted in helium, which would suppress both the helium signal and the corresponding inferred mass-loss rate. \citet{Fossati22} modeled the composition of WASP-80b's outflow and showed that helium depletion (with $n_\mathrm{He}/n_\mathrm{H} \sim 0.01$) might plausibly explain these data. Previous studies of HD 209458b, HD 189733b, and GJ 3470b have sought to quantify the fractional abundances of hydrogen and helium in planetary outflows by comparing the magnitude of absorption measured in both the Lyman-$\alpha$ and metastable helium lines to Parker wind models \citep{Lampon20, Lampon21a, Lampon21b} or 3D hydrodynamical models \citep{Shaikhislamov21}. However, Lyman-$\alpha$ is a poor tracer for the density distribution in the planetary thermosphere, typically probing above the exobase at high velocities \citep[e.g.][]{Owen21}. This suggests that anchoring the Parker wind model composition near the wind-launching radius to models of Lyman-$\alpha$ absorption may be inadvisable. A more reasonable route would be to jointly model H$\alpha$ and metastable helium absorption signals, as suggested by \citet{Czesla22}. Such joint models would be invaluable in resolving this question. 

Strong magnetic fields can also confine planetary outflows. For a strongly ionized wind launched in a dipolar planetary magnetic field, equatorial field lines are closed (ionized material traveling along field lines would loop back to the planet), leading to an equatorial ``dead zone'', while polar field lines remain open \citep{Adams11, Trammell11, Owen14}. For close-in planets orbiting exceptionally active stars, the outflows are largely ionized, so material should follow field lines if the planets are magnetized. WASP-80b and WASP-52b have the smallest semimajor axes and their host stars have the largest $\log(R_\mathrm{HK}')$ values in our whole sample, making this an attractive explanation.

The ratio of magnetic pressure to ram pressure is a good diagnostic for the influence of magnetic fields \citep{Owen14}, but this ratio depends on the magnetic field strength which is quite uncertain. There are several lines of evidence indicating that the magnetic fields of hot Jupiters may be relatively strong (10-100~G), including the inflated radii of these planets \citep{Batygin10, Yadav17} and evidence for magnetic star-planet interactions in a few systems \citep[SPI;][]{Cauley19}. Other observations suggest that these planets may have more modest field strengths (1-10~G), including ultraviolet observations of Lyman-$\alpha$ and ionized carbon lines in HAT-P-11b \citep{BenJaffel22}, and a small inferred hotspot offset in the phase curve of WASP-76b which might be caused by Lorentz drag \citep{May21, Beltz22}. Elsasser number scalings also predict more moderate magnetic field strengths \citep{Stevenson03}. If the Elsasser number is of order unity, the field strength scales with the square root of the rotation rate; for tidally-locked hot Jupiters, this implies $B\sim1/\sqrt{P}$. Assuming the fluid densities and conductivities are similar to that of Jupiter, we have $B\sim B_\mathrm{J}(P/10~\mathrm{hr})^{-1/2}$, of order a few~G for the planets in consideration here. Even at the lower end of this range, magnetic pressure is expected to dominate over ram pressure near the wind-launching region \citep{Owen14}, suggesting that magnetic fields may indeed play a role in confining the outflows of WASP-80b and WASP-52b. 

In summary, inaccuracies in the integrated XUV fluxes are unlikely to explain the helium non-detection for WASP-80b and the small inferred mass-loss rate for WASP-52b, but inaccuracies in the spectral shape could decrease the helium signals for these planets. The stellar wind might also be a contributing factor for WASP-52b, but it is unlikely to explain the non-detection for WASP-80b. Composition and magnetic fields could both play a role in explaining the observations. We propose some observational tests in Section~\ref{sec:conc} that could help to distinguish between these competing hypotheses.

\subsection{The Upper Neptune Desert Is Stable Against Mass Loss}
At the beginning of this work, we hypothesized that if the Neptune desert is indeed cleared by mass loss, then planets at its boundaries should be marginally stable against photoevaporation. We can now test this hypotheses using our measurements and some simple calculations. First, we inspect our observationally-constrained mass-loss rates to evaluate our sample's present-day stability to atmospheric erosion. We expect that close-in giant planets will be destroyed while their host stars are on the main sequence \citep{Hamer19}, so a reasonable timescale for atmospheric stability is $M_\mathrm{env}/\dot{M}\sim4$~Gyr. In Table~\ref{table:massloss}, all of the planets in our sample near the upper edge of the desert have predicted atmospheric lifetimes that are much greater than this timescale. 

For HAT-P-26b near the lower edge of the desert, our result is inconclusive due to the $\dot{M}-T_0$ degeneracy: if the outflow is hot ($T_0\gtrsim~10,000$~K) the planet's predicted atmospheric lifetime may be as short as 3.3~Gyr. This is substantially shorter than the envelope mass-loss timescale for HAT-P-11b: using the mass-loss rate from \citep{dosSantos22} and the envelope mass fraction of 15.1\% from \citep{Lopez14}, the mass-loss timescale for HAT-P-11b is $\sim30$~Gyr. Importantly, the $\dot{M}-T$ degeneracy for HAT-P-11b is broken by the precise line shape measurement \citep{Allart18, dosSantos22} and also by energetic arguments \citep{Vissapragada22}. Both methods suggest the outflow temperature for HAT-P-11b is $T_0 \lesssim 8000$~K, and if this is the case for HAT-P-26b as well, it would have a longer inferred envelope loss timescale. In the future, high-resolution spectroscopy of HAT-P-26b could resolve this degeneracy by measuring the helium line shape and corresponding outflow temperature.

Although the present-day mass-loss rates of the planets in our sample appear to be low, they experienced stronger XUV irradiation (and thus suffered higher mass-loss rates) in the past. We can reconstruct their past mass-loss histories by integrating the energy-limited mass-loss rate back in time \citep[see e.g.][]{Lecavelier07, Jackson12, Mordasini20}. We first start with the expression:
\begin{equation}
    \dot{M} = -\frac{\varepsilon R_\mathrm{p}^3 L_\mathrm{XUV}}{4 a^2 K G M},
\end{equation}
which is the typical expression for the energy-limited mass-loss rate (Equation~\ref{energylim}) with the scaled luminosity $L_\mathrm{XUV}/(4\pi a^2)$ substituted for the flux $F_\mathrm{XUV}$. For this estimate, we assume that the planetary radius is approximately invariant to small changes in the mass. We can do this because the mass-radius relation is nearly flat for the giant planets at the upper edge of the Neptune desert \citep[e.g.][]{ChenKipping17, Owen18, Thorngren19}. This is because giant planets at the upper edge of the desert are well-approximated by $P \propto \rho^2$ polytropes for which the radius does not depend on mass \citep{Stevenson82}. Additionally, the variations in $K$ with mass are also small and can be ignored. The integral then reads:
\begin{equation}
    \int_{M_0}^{M_\mathrm{p}} M dM = \int_0^{t} -\frac{\varepsilon R_\mathrm{p}^3 L_\mathrm{XUV}}{4 a^2 K G} dt,
\end{equation}
where we have introduced the initial planetary mass $M_0$ and the planet's current age $t$. To allow for maximum possible mass loss, we assume that the planet has maintained the same mass-loss efficiency for its whole life, even though the efficiency at early times is likely much lower in the recombination limit \citep{MurrayClay09, Owen16}. Additionally, we assume the planet has remained at its current semimajor axis for its whole life, i.e., that it did not migrate from a more distant location where it experienced lower instellation. Then, defining the integrated stellar XUV luminosity $E_\mathrm{XUV} = \int_0^t L_\mathrm{XUV} dt$, we obtain the equation:
\begin{equation}
    M_\mathrm{p}^2 - M_0^2 = -\frac{\varepsilon R_\mathrm{p}^3E_\mathrm{XUV}}{2a^2KG}.
\end{equation}
Finally, writing the current mass of the planet as a fraction $f < 1$ of the initial mass (such that $M_0 = M_\mathrm{p}/f$), we obtain:
\begin{equation}
    M_\mathrm{p} = \frac{1}{a}\sqrt{\Bigg(-\frac{1}{1 - \frac{1}{f^2}}\Bigg)\frac{\varepsilon R_\mathrm{p}^3E_\mathrm{XUV}}{2KG}}
\end{equation}
This is a line in the $M_\mathrm{p}-a$ plane above which planets have lost less than a fraction $f$ of their initial mass, and below which they have lost more. We hereafter take $f = 0.5$ -- a planet losing half of its initial mass -- as a metric for marginal stability. To calculate this boundary, we require a prescription for the planetary radii and the integrated stellar XUV flux in addition to our empirically-constrained distribution for $\varepsilon$. For the planetary radii, we use the empirical radius-temperature relation from Equation~(1) of \citep{Owen18} for a typical K dwarf temperature $T_\star = 4700$~K and radius $R\star = 0.8R_\Sun$. The spread in the relation is incorporated into our final uncertainties. We note that this relation is only valid for $M_\mathrm{p} \gtrsim 0.2 M_\mathrm{J}$; below this mass the planetary radius will change as the planet undergoes photoevaporation. For the integrated XUV flux, a reasonable upper limit for a K dwarf is $E_\mathrm{XUV}\sim10^{46}$~erg using the fast rotator track from \citet{Johnstone21} as a guide (see their Figure~18). 

\begin{figure*}[ht!]
\centering
\includegraphics[width=\textwidth]{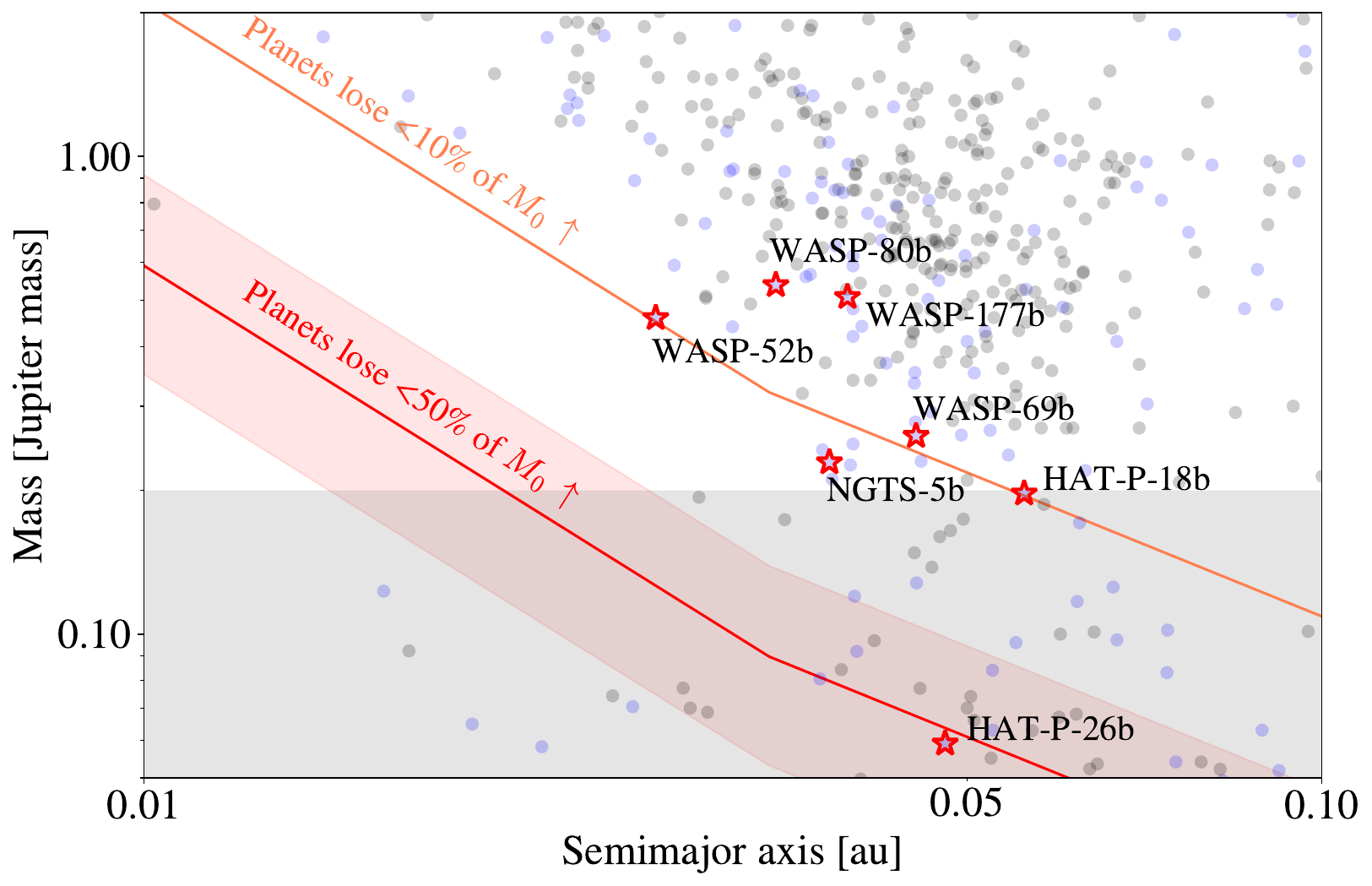}
\caption{$M_\mathrm{p}-a$ plane with marginal stability curves. Planets above the red curve cannot have lost more than 50\% of their initial mass $M_0$ to photoevaporation, even assuming an energy-limited outflow for the entire planetary lifetime. The red shaded region corresponds to the uncertainty on our empirical estimate for the mass-loss efficiency $\varepsilon$ with the uncertainty in the empirical radius-temperature relation from \citet{Owen18} included as well. The orange curve shows the same for 10\% mass loss. Points indicate transiting exoplanets with fractional mass uncertainties of less than 30\%; those orbiting stars with $4000$~K$<T_\mathrm{eff}<5400$~K are colored blue, and the seven targets constituting our sample are labeled and outlined with red stars. The gray shaded region indicates $M_\mathrm{p} < 0.2 M_\mathrm{J}$ wherein our assumed radii are incorrect. }
\label{fig:finalfig}
\end{figure*}

In Figure~\ref{fig:finalfig}, we draw the $f = 0.5$ boundary on the $M_\mathrm{p}-a$ plane for a $R = R_\mathrm{J}$ and $K = 0.5$ planet using our empirical efficiency from Section~\ref{sec:benchmark}. For comparison, we also draw the $f = 0.9$ boundary (corresponding to 10\% mass loss over the planetary lifetime). We conclude that planets at the upper edge of the Neptune desert have lost less than 10\% of their initial masses to photoevaporation. This empirically benchmarked result is in good agreement with previous theoretical calculations by \citet{Ionov18} and \citet{Owen18}. HAT-P-26b, near the lower edge of the Neptune desert, is a potential exception -- it could have lost $\sim50\%$ of its initial mass if its outflow was energy-limited for the entire planetary lifetime. Although planets like HAT-P-26b along the lower edge of the desert could still be substantially affected by mass loss, we caution that our simplifying assumption that the radius is insensitive to small changes in envelope mass will not hold true for these smaller planets, and a more detailed framework taking into account the core-envelope structure is needed for understanding this part of the population \citep[e.g.][]{Owen18, Mordasini20}.

Throughout this section, we made a number of assumptions to try to estimate the maximum possible mass-loss endured by planets in and near the upper edge of the Neptune desert. In reality, not all planetary host stars will have XUV luminosities as large as those of the most rapidly rotating stellar models from \citet{Johnstone21}. We also assumed an energy-limited outflow for the entire planetary lifetime; this is certainly an overestimate at early times, and the efficiency will be much lower during the period of maximum XUV irradiation \citep{MurrayClay09, Owen16}. Despite this, we still found that the boundary for marginal stability is located well below the actual edge of the desert. We conclude that photoevaporation cannot explain the upper boundary of the Neptune desert.

\section{Conclusions} \label{sec:conc}
In this work, we searched for helium outflows in 7 gas-giant exoplanets orbiting K-type host stars over the course of a 12-night survey with Palomar/WIRC. We summarize our results below:

\begin{enumerate}
    \item We detected ($>3\sigma$ confidence) helium absorption from WASP-69b, HAT-P-18b, and HAT-P-26b; tentatively detected ($2-3\sigma$) absorption from WASP-52b and NGTS-5b; and did not detect ($<2\sigma$) absorption from WASP-80b and WASP-177b.
    \item When interpreting these signals with a one-dimensional Parker wind model, we found that the six planets in our sample near the upper edge of the desert (WASP-69b, WASP-52b, HAT-P-18b, WASP-80b, WASP-177b, and NGTS-5b) have atmospheric lifetimes $M_\mathrm{env}/\dot{M}$ much larger than 10~Gyr. Our result for HAT-P-26b (near the lower edge of the desert) is inconclusive due to the $\dot{M}-T_0$ degeneracy.
    \item We compared our empirically measured mass-loss rates to predictions from the one-dimensional, self-consistent hydrodynamics code ATES \citep{Caldiroli21, Caldiroli22}. We found that five planets were in good agreement with the ATES predictions: HAT-P-18b, WASP-69b, HAT-P-26b, NGTS-5b, and WASP-177b. The mass-loss rates for these planets are all consistent with a mean outflow efficiency of $\varepsilon = 0.41^{+0.16}_{-0.13}$.
    \item We found that WASP-52b and WASP-80b have much lower inferred mass-loss rates than predicted by these models. The measured helium absorption signals for these two planets may be affected by stellar wind confinement, helium depletion, the inverse first ionization potential effect, or magnetic field confinement, although stellar wind confinement appears unlikely for WASP-80b.
    \item Our empirically measured mass-loss efficiencies are too small for photoevaporation to sculpt the population of giant planets at the upper edge of the Neptune desert, in agreement with previous work by \citet{Owen18} and \citet{Ionov18}.

\end{enumerate}

We conclude that another mechanism besides photoevaporation must be responsible for carving the upper edge of the Neptune desert. Candidate explanations include high-eccentricity migration \citep{Matsakos16, Owen18} and \textit{in situ} formation near the magnetospheric truncation radius of the natal protoplanetary disk \citep{Bailey18}. Although our observations cannot distinguish between these two scenarios, future studies of the functional dependence of the desert's upper boundary on stellar properties may yield more insight into the problem. This will require a large, statistically uniform sample of hot Jupiters, which will soon be provided by \textit{TESS} \citep{Yee21, Yee22}. Additional observations aimed at constraining the presence of long-period planetary companions, for instance with \textit{Gaia} astrometry, could also help to differentiate between these hypotheses. 

Our survey also revealed that 1D mass loss models overpredict the metastable helium signatures from WASP-52b and WASP-80b, in agreement with previous work by \citet{Fossati22}. Although it cannot explain WASP-80b, our estimates suggest that a strong stellar wind might confine the outflow of WASP-52b, similar to what has been proposed for WASP-107b \citep{Spake21, Wang21b} and TOI-560.01 \citep{Zhang22}. If the helium line for WASP-52b can be detected with more sensitive observations at higher resolving power, we would expect to see a characteristic blueshifted line profile and/or an extended tail of post-egress absorption in this scenario. If the outflow is instead confined by magnetic fields, the line profile would not be strongly Doppler shifted. We could also test the magnetic confinement explanation for both planets by searching for evidence of star-planet interactions. \citet{Cauley19} recently demonstrated that close-in planets with strong magnetic fields can affect the stellar chromospheric emission. These magnetic star-planet interactions (SPI) are detected in the Ca II K line, where variations in stellar emission occur on the planetary orbital timescale. It is also possible that the stellar XUV templates we have used for these stars are inadequate as the coronal iron abundances may differ between WASP-80/WASP-52 and the template star $\epsilon$ Eridani \citep{Poppenhaeger22}; this motivates future X-ray monitoring of these sources. Lastly, observations of the H$\alpha$ absorption signal from both planets could also provide constraints on the outflow composition \citep{Czesla22}. H$\alpha$ absorption has already been detected and used to constrain composition for WASP-52b \citep{Chen20, Yan22}; similar constraints should also be obtained for WASP-80b.

During the review process for this paper, \citet{Kirk22} announced a strong detection of helium absorption in WASP-52b and tentative evidence for helium absorption in WASP-177b, both with Keck/NIRSPEC. For WASP-177b, their result is consistent with our upper limit, but is more constraining on the mass-loss rate: they obtain a $3\sigma$ upper limit of $7.9\times10^{10}$~g~s$^{-1}$ on the mass-loss rate. The \citet{Kirk22} result matches the posterior distribution shown in Figure~\ref{fig:caldiroliplot} quite well after accounting for the $K = 0.52$ correction factor. For WASP-52b, \citet{Kirk22} observe a slightly stronger signal than ours: 0.66$\pm$0.14\% (in a 0.635~nm bin) compared to our result of 0.29$\pm$0.13\%, and they obtain a mass-loss rate of $1.2\pm0.5\times10^{11}$~g~s$^{-1}$. This is still an order of magnitude smaller than the prediction from ATES in Figure~\ref{fig:caldiroliplot} ($K\dot{M} = 4.7\times10^{11}$~g~s$^{-1}$) after accounting for the $K = 0.43$ correction factor. Both \citet{Kirk22} and the subsequent modeling effort by \citet{Yan22} find similar inferred outflow rates that are not very sensitive to the assumed composition. Additionally \citet{Kirk22} find that the outflow is not strongly confined by stellar winds as they do not observe a Doppler-shifted line profile or a tail. Of the factors we considered for the low inferred mass-loss rate for WASP-52b compared to model predictions, this leaves the inverse first ionization potential effect and/or confinement by magnetic fields as the most likely explanations.

Intriguingly, the \textit{TESS} survey is also beginning to discover planets \textit{within} the Neptune desert, including TOI-849b \citep{Armstrong20} and LTT 9779b \citep{Jenkins20}. \citet{Dai21} show that these desert-dwellers tend to orbit metal-rich stars and lack planetary companions, suggesting that planets within the desert are more similar to gas giant planets than the rocky, ultra-short period planets below the desert. However, our result confirm that these desert-dwellers are unlikely to be the photoevaporated cores of more massive gas giant planets. In the high-eccentricity migration scenario, these unique planets could be the results of partial tidal disruption as they circularized onto their current orbits \citep{Faber05, Guillochon11}. Alternatively, \citet{Pezzotti21} suggest that if TOI-849b had a large initial mass and radius, it may instead have lost its envelope to Roche-lobe overflow (RLO). This process can lead to mass-loss rates far exceeding those from photoevaporation, and is predicted to be consequential for planets on exceptionally short-period orbits \citep{Jackson17}. While it likely plays a role for the closest-in Neptune desert planets, the RLO mass-loss rate drops off precipitously with orbital distance, so it cannot explain the entirety of the desert \citep{Koskinen22}.

Although these high density desert-dwelling Neptunes likely have very low present-day mass-loss rates, the picture for lower density Neptunes at the lower edge of the desert is somewhat murkier. Previous modeling studies \citep{Kurokawa14, Lundkvist16, Ionov18, Owen18} concluded that mass loss can be significant for planets near the lower edge of the desert. Our result for HAT-P-26b appears to be consistent with this prediction. \textit{TESS} has identified a large sample of new planets near the lower edge of the desert that are favorable targets for mass-loss measurements, making this a promising area for future investigation. If planetary evolution is dominated by photoevaporation in this mass regime, we also expect the lower boundary of the desert to shift to smaller insolations for less luminous late-type stars \citep[e.g.][]{McDonald19, Kanodia21}. This appears to be confirmed observationally \citep{Petigura22}, constituting additional evidence for the importance of mass loss in sculpting the lower part of the desert. With the continued success of \textit{TESS} and the ability to probe mass-loss rates with metastable helium, we now have the means to unveil the divergent evolutionary pathways of planets on either side of the Neptune desert.
    
\acknowledgments
We thank the Palomar Observatory telescope operators, support astronomers, and directorate for their support of this work, especially Tom Barlow, Andy Boden, Carolyn Heffner, Paul Nied, Joel Pearman, Kajse Peffer, and Kevin Rykoski. We also thank Aaron Bello-Arufe, Yayaati Chachan, Philipp Eigm\"{u}ller, Akash Gupta, Julie Inglis, Shubham Kanodia, Katja Poppenhaeger, Hannah Wakeford, and Nicole Wallack for insightful conversations. We acknowledge the referees for thorough reports that improved the quality of this paper. SV is supported by an NSF Graduate Research Fellowship. HAK acknowledges support from NSF CAREER grant 1555095. AO gratefully acknowledges support from the Dutch Research Council NWO Veni grant.

\facilities{ADS, NASA Exoplanet Archive, Hale 200-inch, TESS}
\software{\texttt{numpy} \citep{Harris20},
\texttt{scipy} \citep{Virtanen20},
\texttt{astropy} \citep{Astropy13, Astropy18},
\texttt{matplotlib} \citep{Hunter07},
\texttt{p-winds} \citep{dosSantos22},
\texttt{theano} \citep{Theano16}, \texttt{pymc3} \citep{Salvatier16}, \texttt{exoplanet} \citep{ForemanMackey21a, ForemanMackey21b}, \texttt{ldtk} \citep{Husser13, Parviainen15}, \texttt{corner} \citep{ForemanMackey16}} 

\clearpage

\appendix 

\section{Limb Darkening Coefficients} \label{ldcs}
\begin{figure*}[h]
\centering
\includegraphics[width=\textwidth]{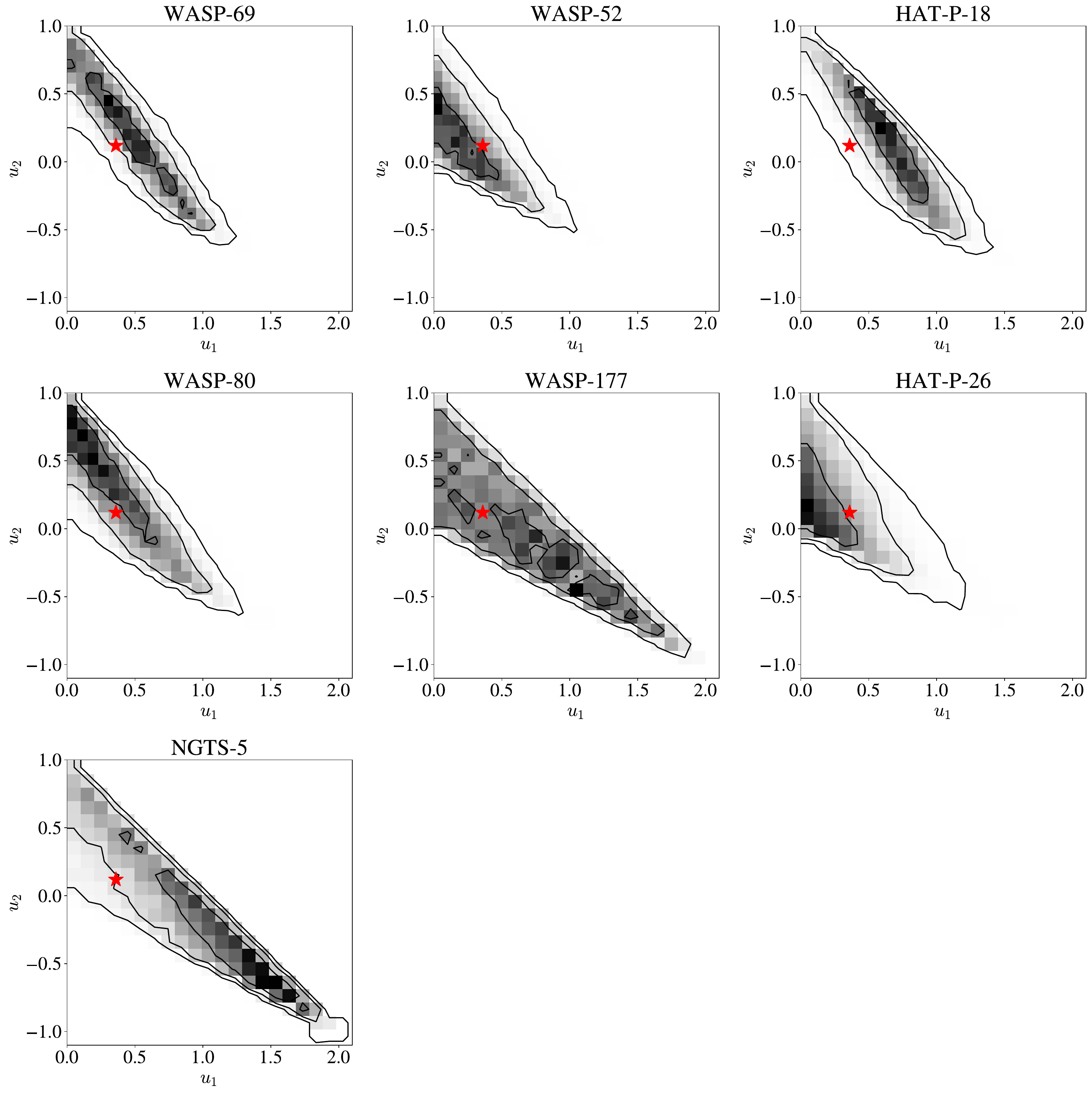}
\caption{Posteriors on the quadratic limb darkening coefficients for all stars analyzed in this work. Contours indicate the 1, 2, and 3$\sigma$ levels on the posterior mass. The red star on each plot indicates the limb darkening coefficient in the helium bandpass computed using \texttt{ldtk} \citep{Husser13, Parviainen15}.}
\label{fig:ldcs}
\end{figure*}

In Figure~\ref{fig:ldcs}, we present the posteriors on the limb darkening coefficients for all stars analyzed in this work alongside calculations for the coefficients with \texttt{ldtk} \citep{Husser13, Parviainen15}. 

\section{Detrending Vector Selection} \label{modelselection}
In Table~\ref{table:detrend}, we summarize the $\Delta$BIC values obtained when comparing models with different combinations of detrending vectors.

\begin{deluxetable*}{cccccccccc}[h]
\tablecaption{Summary of $\Delta$BIC between detrending models in fits with free limb-darkening coefficients \label{table:detrend}.}
\tablehead{\colhead{Planet}  & \colhead{Date}  & \colhead{LDC} & \colhead{1} & \colhead{2} & \colhead{3} & \colhead{1,2} & \colhead{1,3} & \colhead{2,3} & \colhead{1,2,3}}
\startdata
WASP-69b & 2019 Aug 16 & free & 6.27 & 0.13 & -10.11 & 5.69 & -5.19 & \textbf{-10.22} & -3.61 \\
 &  & \textbf{fixed} & 7.18 & 1.04 & \textbf{-9.98} & 6.09 & -3.39 & -8.44 & -3.30 \\
WASP-52b & 2019 Sep 17 & free &-5.68 & 2.58 & -0.06 & -6.62 & \textbf{-10.93} & 3.51 & -5.30 \\
&  & \textbf{fixed} &-4.23 & 4.10 & 2.26 & -3.75 & \textbf{-8.22} & 4.75 & -3.49 \\
HAT-P-18b & 2020 Jun 05 & \textbf{free} & 5.49 & 3.47 & 3.62 & 8.70 & 8.64 & 7.83 & 14.02 \\
 & & fixed & 4.66 & 1.99 & 3.15 & 7.97 & 9.51 & 6.86 & 13.79 \\
HAT-P-18b & 2020 Jul 08 & \textbf{free} & \textbf{-1.26} & 1.09 & 3.62 & 0.25 & 3.56 & 7.65 & 7.10  \\
&  & fixed & \textbf{-0.42} & 1.15 & 3.43 & 1.90 & 3.83 & 6.65 & 7.40  \\
WASP-80b & 2020 Jul 09 (first half) & free & 5.43 & -23.35 & 2.26 & -19.19 & 7.65 & \textbf{-34.50} & -28.68 \\
&  & \textbf{fixed} & 4.76 & -22.83 & 2.73 & -17.00 & 7.35 & \textbf{-31.85} & -26.11 \\
WASP-80b & 2020 Jul 09 (second half) & free &3.21 & 2.41 & 3.39 & 7.00 & 6.07 & 7.80 & 11.96 \\
& & \textbf{fixed} & 3.52 & 3.46 & 2.58 & 7.59 & 7.48 & 8.01 & 12.14 \\
WASP-52b & 2020 Aug 04 & free & 5.55 & \textbf{-7.27} & -4.70 & -2.55 & 0.31 & -2.43 & 4.37 \\
& & \textbf{fixed} & 5.04 & \textbf{-8.49} & -5.43 & -3.01 & -0.02 & -2.58 & 2.68 \\
WASP-177b & 2020 Oct 02 & free & 5.06 & 5.38 & 5.68 & 9.67 & 10.07 & 7.98 & 11.76 \\
& & \textbf{fixed} & 4.94 & 4.57 & 4.92 & 8.97 & 9.85 & 7.24 & 12.80 \\
HAT-P-26b & 2021 May 01 & free & 1.19 & \textbf{-2.90} & 7.10 & 0.57 & 6.89 & 0.15 & 3.31 \\
&  & \textbf{fixed} & -0.43 & \textbf{-5.35} & 5.60 & -1.79 & 5.07 & -1.38 & 1.40 \\
HAT-P-26b & 2021 May 18 & free & 3.54 & 5.80 & 0.77 & 10.47 & 5.60 & \textbf{-9.22} & -4.17 \\
 & & \textbf{fixed} & 4.53 & 5.57 & 0.87 & 9.92 & 6.15 & \textbf{-8.56} & -2.91 \\
NGTS-5b & 2021 May 27 & free & 4.26 & \textbf{-26.18} & -26.18 & -23.46 & -22.01 & -21.70 & -18.07 \\
&  & \textbf{fixed} & 5.29 & \textbf{-26.87} & -25.87 & -22.53 & -21.66 & -21.31 & -18.08 \\
\enddata
\tablecomments{All $\Delta$BIC values are relative to that for no additional detrending vectors, with negative values indicating favored models. Vector 1 corresponds to the centroid offsets, vector 2 corresponds to the water absorption proxy, and vector 3 corresponds to the airmass. Bolded values indicate the adopted detrending model for a given planet; if there are no bolded values then the model with no additional detrending vectors was selected. The ``LDC" column denotes the fitting strategy of leaving the limb darkening coefficients free or fixing them to a calculation from \texttt{ldtk} \citep{Husser13, Parviainen15}; the adopted strategy for each planet is noted in bold.}
\end{deluxetable*}

\section{Allan Deviation Plots} \label{allan}
\begin{figure*}[h]
\centering
\includegraphics[width=\textwidth]{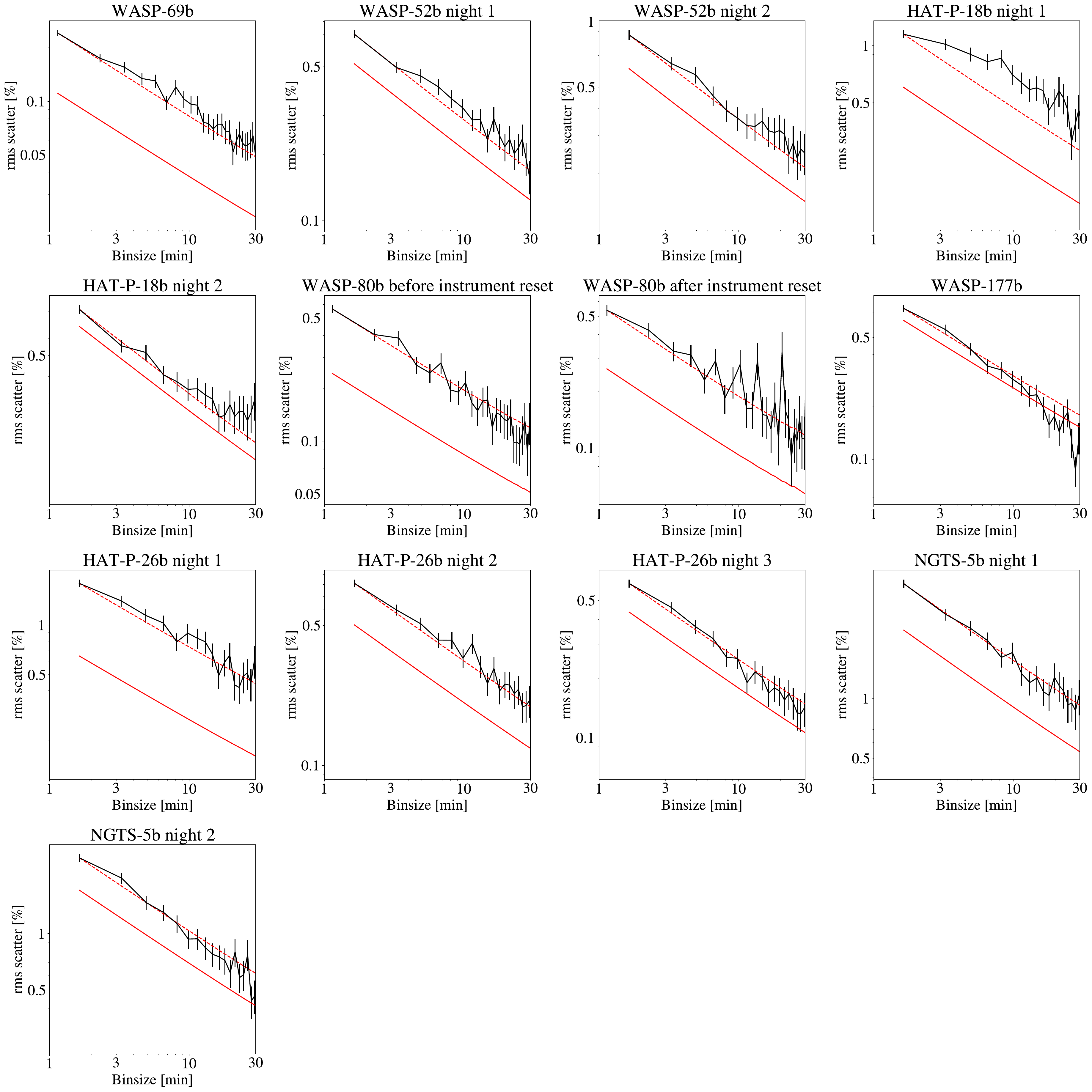}
\caption{Allan deviation plots for all light curves analyzed in this work. The rms of the binned residuals for each light-curve fit are shown with the black curves, the red noise indicates the expectation from photon noise statistics, and the dashed red line is the red line scaled up to match the first point of the black curve. The photon noise values and scaling factors are given in Table~\ref{table:log}.}
\label{fig:allan}
\end{figure*}

In Figure~\ref{fig:allan}, we present the Allan deviation plots for all light curves analyzed in this work.

\clearpage
\section{\textit{TESS} Light-Curve Fit} \label{tessfit}
\begin{figure}[h]
\centering
\includegraphics[width=0.45\textwidth]{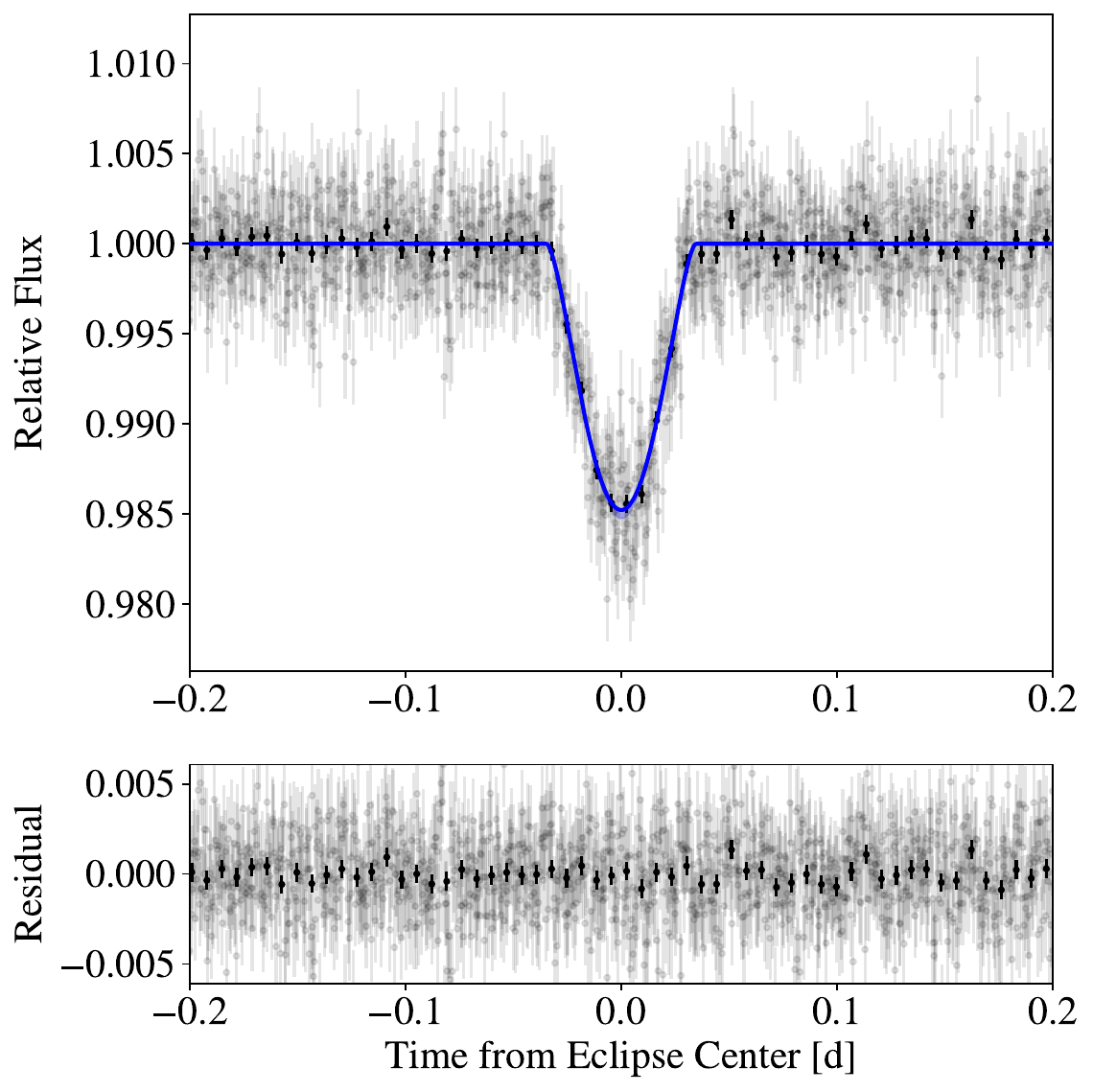}
\caption{Same as Figure~\ref{fig:w69}, but for the WASP-177b \textit{TESS} data.}
\label{fig:w177tess}
\end{figure}

In Figure~\ref{fig:w177tess}, we present the light-curve fit for the WASP-177b \textit{TESS} data. We found a radius ratio in the \textit{TESS} bandpass of $0.187_{-0.036}^{+0.054}$. The rest of the fit parameters are given in Table~\ref{table:posteriors} as this was a joint fit with the WIRC data.

\label{tess}

\section{The Parker Wind Model with Tidal Gravity} \label{tidal}
We demonstrate that a one-dimensional transonic Parker wind model should achieve the sound speed:
\begin{equation}
    c_\mathrm{s} = \sqrt{\frac{k_\mathrm{B}T_0}{\bar{\mu}m_\mathrm{p}}}
\end{equation}
within the planetary Roche lobe ($R_\mathrm{s} < R_\mathrm{Roche}$) when the tidal gravity term is considered \citep{MurrayClay09}. Here, $M_\mathrm{p}$ is the planet mass, $G$ is the gravitational constant, $k_\mathrm{B}$ is the Boltzmann constant, $T_0$ is the isothermal outflow temperature, $\bar{\mu}$ is the average molecular weight as defined by \citet{Lampon20}, and $m_\mathrm{p}$ is the proton mass. The sonic point without considering the stellar gravity is defined as: 
\begin{equation}
    R_\mathrm{s} = \frac{GM_\mathrm{p}}{2c_\mathrm{s}^2} = \frac{GM\bar{\mu} m_\mathrm{p}}{2k_\mathrm{B}T_0}. \label{sonic}
\end{equation}

To derive the sonic point with tidal gravity considerations, we first write the full momentum equation for the wind \citep[e.g. Equation (2) from][]{MurrayClay09}: 
\begin{equation}
    v\frac{dv}{dr} + \frac{1}{\rho}\frac{dP}{dr} + \frac{GM_\mathrm{p}}{r^2} - \frac{3GM_\star r}{a^3} = 0.
\end{equation}
As in the original Parker wind solution, we rewrite the pressure gradient term with a density gradient using the derivative of the ideal gas law. We then rewrite the density gradient term as a velocity gradient term using the derivative of the continuity equation \citep[a step-by-step walkthrough of this method is provided in Chapter 3 of][]{Lamers99}. The result is:
\begin{equation}
\frac{1}{v}\frac{dv}{dr} = \frac{1}{v^2 - c_\mathrm{s}^2}\Big(\frac{2c_\mathrm{s}}{r} - \frac{GM_\mathrm{p}}{r^2} + \frac{3GM_\star r}{a^3} \Big).
\label{newparker}
\end{equation}
The sonic point is the singular point of Equation~(\ref{newparker}) where the numerator and denominator of the right-hand side both go to zero. The latter condition gives $v = c_\mathrm{s}$, identically to the original Parker wind solution. Requiring the numerator to go to zero gives the cubic:
\begin{equation}
    \frac{3GM_\star r^3}{a^3} + 2c_\mathrm{s}^2r - GM_\mathrm{p} = 0.
\end{equation}
This is the major difference from the Parker wind considered by \citet{Oklopcic18}. Neglecting the tidal gravity term, we obtain the original sonic point solution in Equation~(\ref{sonic}), but the general solution behaves differently:
\begin{equation}
    R_\mathrm{s} = a\Bigg(\sqrt[3]{\frac{M_1 + \frac{M_\mathrm{p}}{2}}{3M_\star}} - \sqrt[3]{\frac{M_1 - \frac{M_\mathrm{p}}{2}}{3M_\star}}\Bigg), \label{newsonic}
\end{equation}
where $M_1$ is further defined as:
\begin{equation}
    M_1 = \sqrt{\Big(\frac{M_\mathrm{p}}{2}\Big)^2 + \frac{8M_\star^2}{81}\Big(\frac{c_\mathrm{s}}{v_\mathrm{K}}\Big)^6}, \label{m1}
\end{equation}
and $v_\mathrm{K} = \sqrt{GM_\star/a}$ is the Keplerian velocity. Equation~(\ref{newsonic}) agrees with Equation~(\ref{sonic}) in the limit of large sound speed, but the crucial point here is to consider the case where the sound speed is very small, corresponding to cold outflows. Whereas the original sonic point solution would have these outflows fall outside the Roche radius, in this limit $M_1 = M_\mathrm{p}/2$, and substitution into Equation~(\ref{newsonic}) yields $R_\mathrm{s} = R_\mathrm{Roche}$. A similar result is obtained in Appendix~C of \citet{Tang20}.

Therefore, when the sonic point in our original models (which neglected the tidal gravity term) began to venture outside the Roche lobe, it meant that our model was not providing an adequate representation of the density and velocity profiles, which are dominated by the stellar gravity. In general, Equation~(\ref{m1}) shows that stellar gravity begins to dominate the outflow behavior when:
\begin{equation}
    \frac{32}{81}\Big(\frac{c_\mathrm{s}}{v_\mathrm{K}}\Big)^6\Big(\frac{M_\star}{M_\mathrm{p}}\Big)^2 \lesssim 1. \label{breakdown}
\end{equation}
Expanding out the Keplerian velocity term on the left-hand side, this condition scales with $c_\mathrm{s}^6a^3/(M_\star M_\mathrm{p}^2)$; therefore the tidal gravity term is important in planets with cool outflows and planets in close proximity to more massive stars. For our sample, the relevant quantities are approximately  $v_\mathrm{K}\sim100$~km/s, $c_\mathrm{s}\sim10$~km/s, and $M_\mathrm{p}/M_\star \sim 3\times10^{-4}$, and Equation~(\ref{breakdown}) is nearly satisfied, indicating that the tidal gravity term in the momentum equation appreciably alters our results. The new momentum equation (Equation \ref{newparker}) is readily integrated from the sonic point $(R_\mathrm{s}, c_\mathrm{s})$ to an arbitrary point $(r, v(r))$ to obtain the velocity profile:
\begin{align}
    \frac{v(r)}{c_\mathrm{s}}\exp{\Big(-\frac{v(r)^2}{2c_\mathrm{s}^2}\Big)} = &\Big(\frac{R_\mathrm{s}}{r}\Big)^2\exp\Bigg[-\frac{GM_\mathrm{p}}{c_\mathrm{s}^2r} + \frac{GM_\mathrm{p}}{c_\mathrm{s}^2R_\mathrm{s}} \nonumber \\
     &- \frac{3GM_\star r^2}{2a^3c_\mathrm{s}^2} + \frac{3GM_\star R_\mathrm{s}^2}{2a^3c_\mathrm{s}^2} - \frac{1}{2}
    \Bigg],
\end{align}
and using the continuity equation we can obtain the density profile:
\begin{align}
    \frac{\rho(r)}{\rho_\mathrm{s}} = &\exp\Bigg[\frac{GM_\mathrm{p}}{c_\mathrm{s}^2r} - \frac{GM_\mathrm{p}}{c_\mathrm{s}^2R_\mathrm{s}} \nonumber \\
     &+ \frac{3GM_\star r^2}{2a^3c_\mathrm{s}^2} - \frac{3GM_\star R_\mathrm{s}^2}{2a^3c_\mathrm{s}^2} + \frac{1}{2} - \frac{v(r)^2}{2c_\mathrm{s}^2}
    \Bigg].
\end{align}
The updated Parker wind structure has been included into the latest release of the \texttt{p-winds} code \citep{dosSantos22}.

\end{document}